\documentclass[%
aip,
jcp,
reprint, 
amsmath,amssymb,
]{revtex4-2}

\usepackage{graphicx}
\usepackage{dcolumn}
\usepackage{bm}

\usepackage[utf8]{inputenc}
\usepackage[T1]{fontenc}
\usepackage{mathptmx}
\usepackage{etoolbox}
\usepackage[version=4]{mhchem}
\usepackage{soul}
\usepackage{siunitx}
\usepackage{booktabs}
\usepackage{tabularx}

\makeatletter
\def\@email#1#2{%
	\endgroup
	\patchcmd{\titleblock@produce}
	{\frontmatter@RRAPformat}
	{\frontmatter@RRAPformat{\produce@RRAP{*#1\href{mailto:#2}{#2}}}\frontmatter@RRAPformat}
	{}{}
}%
\makeatother

\begin{document}
	
	\preprint{AIP/123-QED}
	
	\title{Beyond Gaussian Statistics in Polymer Melts: Statistical Masking of Persistent Local Constraints}
    \author{José  A. Martins}
	\affiliation{Departamento de Engenharia de Polímeros, Universidade do Minho, Campus de Azurém, 4800-058 Guimarães, Portugal} 
	\email{jamartinspessoal@gmail.com}
	
	\date{\today}%
	
	%
	
\begin{abstract}
	Short polymer chains exhibit clear deviations from Gaussian end-to-end distance statistics, yet the molecular mechanism by which Gaussian behavior is recovered in long chains remains unestablished.
	Atomistic molecular dynamics simulations of polyethylene melts reveal that conformational heterogeneity persists at the Kuhn scale across all chain lengths, consisting of a mosaic of slow-relaxing, extended aligned chain segments (ACS) and coiled segments --- random conformational sequences (RCS) and chain ends (CE).
	We show that the end-to-end distance distributions for both unentangled and entangled chains are accurately described by a $q$-Gaussian function, with the entropic index $q$ increasing systematically from $0.67$ (C50) to $0.99$ (C500). 
	This evolution tracks the emergence and accumulation of RCS segments, which are absent in short chains, establishing $q$ as a quantitative ``heterogeneity index''. 
	The $q < 1$ values are a signature of non-extensive statistics, with the ratio of Tsallis to Boltzmann-Gibbs entropy ($S_q/S_1$), computed directly from simulation data without fitting, decreasing from $1.80$ (C50) to $1.03$ (C500).
	Crucially, we demonstrate that Gaussian recovery does not result from the erasure of Kuhn-scale heterogeneities, as ACS domains persist in all chain lengths above the critical mass ($\approx 35\%$). Instead, the transition to Gaussian statistics is a statistical masking effect, where the accumulation of independent RCS segments progressively obscures the non-Gaussian signatures of the persistent ACS domains.
\end{abstract}
	
\maketitle

\section{\label{sec:introduction}INTRODUCTION}

The classical description of polymer chains as ideal Gaussian coils is a cornerstone of polymer physics. 
Based on the random-walk analogy, this model assumes identical, independent statistical segments and additive (extensive) entropy. Under these conditions,
the end-to-end vector distribution is Gaussian, and many macroscopic
properties can be predicted.\cite{Rubinstein-Colby-PolPhys-2003,	Lodge-Heimenz-PolyChem-2007,Gedde-2019,Teraoka-PolymSol-2002}

However, real polymer chains exhibit intrinsic conformational heterogeneity at the minimal statistical segment level, the Kuhn segment. 
Conformational heterogeneity in macromolecular systems is supported, beyond any reasonable doubt, by an extensive set of experimental and simulation results.

As early as 1991, \citeauthor{Frauenfelder-1991} established for proteins that conformational substates with distinct structural and dynamic properties exist and are organized into a coarse hierarchy,\cite{Frauenfelder-1991} a picture that finds its direct analogue in the coexistence of aligned chain segments and random
conformational sequences at the Kuhn scale in polymer
melts.\cite{Martins-macromol-2013,Martins-2026}

For synthetic polymers, on the experimental side, time-resolved FTIR and SAXS measurements by \citeauthor{Tashiro-1998} and \citeauthor{Tashiro-1999} on polyethylene revealed the formation of conformationally ordered, \textit{trans}-rich sequences during crystallization from the melt, demonstrating that locally extended
backbone segments coexist with disordered regions even in the early stages of ordering from the melt.\cite{Tashiro-1998,Tashiro-1999}
Using polarized Raman spectroscopy, \citeauthor{Migler-trans-rich-2015}
directly identified \textit{trans}-rich structures in polyethylene melts prior to crystallization, confirming that such ordered conformational domains are intrinsic to the melt state rather than a consequence of crystallization.\cite{Migler-trans-rich-2015}
Importantly, analogous regular conformational sequences have been reported in non-crystalline atactic polystyrene gels,\cite{Tashiro-2000} indicating that Kuhn-scale heterogeneity is a generic feature of flexible polymer melts, not a peculiarity of polyethylene chemistry or stereochemical regularity.

On the simulation side, atomistic molecular dynamics studies of polyethylene melts have consistently identified locally ordered, \textit{trans}-rich backbone segments coexisting with disordered regions, across a range of chain lengths and temperatures.\cite{Sundararajan-Kavassalis-JCSoc-1995,Kavassalis-Sundararajan-macrom-1993, Rigby-Roe-1995,Rutledge-2003,Martins-macromol-2013}
These substructures influence crystallization kinetics, local mechanical response, and segmental relaxation. 

In a previous work,\cite{Martins-2026} besides defining the Kuhn segment as the minimal statistical segment, we showed that polymer chains are better described by a 3D mosaic of different types of Kuhn segments:
(i)~aligned chain segments (ACS) with high local alignment;
(ii)~random conformational sequences (RCS), i.e., segments between
two ACS with enhanced conformational freedom and large-amplitude
angular fluctuations; and
(iii)~chain ends (CE) located between an ACS and the chain end, behaving similarly to the RCS.
ACS exhibit extended orientational correlations and relax an order of magnitude more slowly than RCS/CE segments, with stretched-exponential relaxation $\beta_{\text{KWW}}\approx0.5$.
This slow relaxation was explained by combining the works of Skolnick and Helfand,\cite{Skolnick-Helfand-1980} Boyd et~al.,\cite{Boyd-JCP-1994} and Shlesinger and Montroll\cite{Shlesinger-Montroll-1984} with our own results.\cite{Martins-2026} Taken together, these results support a mosaic picture of the polymer melt at the Kuhn scale, in which the chain contour is segmented into dynamically and structurally distinct domains.

Conformational heterogeneities in polymer melts are closely related to non-Gaussianity in polymer systems. In particular, this non-Gaussianity has been studied in the context of dynamic heterogeneities near the glass transition\cite{Ediger-annualreviews-2000,Ediger-2003,Glotzer-PR-E-2001,Glotzer-PRL-1997,Donati-Douglas-PRL-1998} and in entangled melts.\cite{Kruteva-Zamponi-macom-2021,Zamponi-2021,Guenza-2001}
At a more molecular scale, \citeauthor{Colmenero-2011} reported that non-Gaussianity arises from increased intramolecular barriers in short bead--spring chains.\cite{Colmenero-2011}

In the context of heterogeneous soft-matter and biological systems,  \citeauthor{Spakowitz-2017}\cite{Spakowitz-2017} describe several experimental systems where particle diffusion exhibits non-Gaussian displacement behavior. 
They showed that RNA-protein particles in the cytoplasm of living cells exhibit non-Gaussian displacement distributions with subdiffusive motion, attributing the
non-Gaussianity to the combination of heterogeneity between particle trajectories and dynamic heterogeneity within single trajectories.

None of the above works, however, connects non-Gaussianity to non-extensivity. Experimental results for cell motion in aggregates, specifically their velocity distributions,\cite{UPADHYAYA2001549} provided this connection. The velocity distributions of cells were analyzed with both the Maxwell--Boltzmann distribution and the $q$-exponential distribution of velocities; the entropic index $q$
quantified the non-extensivity. 
It was found that cell motion in aggregates was characterized by regimes in which the cell is either trapped in a cage of its fluctuating nearest neighbors, or has a
finite probability of sudden escape from the cage. We note that this conclusion is consistent with a continuous-time random walk dynamics, with pulses and pauses, and subdiffusive motion.\cite{Shlesinger-Montroll-1984}

In the context of polymer simulations specifically, \citeauthor{Ma-Hu-2010-I} showed that the velocity distributions of repeat units in non-equilibrium bead-spring polymer chains are well described by the Tsallis $q$-exponential distribution of velocities, with the entropic index $q$ increasing monotonically with chain
stiffness and converging to unity as constraints weaken.\cite{Ma-Hu-2010-I,Ma-Hu-2010-II,Ma-Hu-2017}

Both \citeauthor{UPADHYAYA2001549} and \citeauthor{Ma-Hu-2017} quantified non-extensivity through the entropic index $q$ obtained from fits of the $q$-exponential distribution of velocities to their data.\cite{UPADHYAYA2001549,Ma-Hu-2017} 
To the best of our knowledge, quantitative evaluations of entropy in known non-Gaussian polymer systems have not yet been performed.

The above results on heterogeneity at the Kuhn scale, non-Gaussianity, and non-extensivity raise several questions.
Are these three features related? 
Is there a molecular origin for the observed non-Gaussianity, and can non-Gaussian behavior in polymer melts be modeled in a unified way? 
Is the crossover to Gaussianity defined only by chains at the critical molar mass $M_c$, without any role of heterogeneities at the Kuhn-segment scale?

The classic SANS experiments of the 1970s by Benoit \textit{et} \textit{al.}\cite{Benoit-1973,Benoit-1974} and Schelten \textit{et} \textit{al.}\cite{Schelten-1974} established, first for polystyrene melts and then for polyethylene melts, that $R_g \propto M^{0.5}$, where $R_g$ is the radius of gyration, consistent with Flory's prediction that chain swelling due to excluded volume is screened by surrounding chains. 
Does this result imply that heterogeneities at the Kuhn scale are erased as the chain reaches Gaussianity? 
As noted above, there exists an extensive body of experimental evidence demonstrating the persistence of conformational heterogeneities in polymer melts.
Why do these heterogeneities apparently not affect the scaling of $R_g$
with molar mass? 
Crucially, SANS measures $R_g$, a single  ensemble-averaged quantity sensitive only to the second moment of $P(R)$, and is therefore insensitive to the shape of the full
end-to-end distance distribution or to subchain heterogeneity at the Kuhn scale.

Finally, since Gaussianity assumes Boltzmann--Gibbs (BG) statistics and entropy extensivity,\cite{Tsallis-2009,Umarov-Tsallis-AIP-2007} can we detect a breakdown of extensivity in systems exhibiting strong deviations from Gaussianity?

To address these questions, we adopt the Tsallis entropy framework\cite{Tsallis-1998,Gell-Mann-Tsallis-2003,Tsallis-2009} and the associated $q$-Gaussian distribution as phenomenological generalizations of BG statistics and the Gaussian distribution, respectively. 
The $q$-Gaussian reduces to the standard Gaussian when $q=1$; values of $q\neq1$ signal non-extensivity. 
The adoption of the $q$-Gaussian is supported by two independent lines of evidence
beyond fit quality: the systematic variation of $q$ with chain length tracks the structural evolution from ACS-dominated to RCS-dominated chain contours, and the ratio $S_q/S_1$, computed directly from simulation data without any fitting, decreases monotonically from $1.80$ (C50) to $1.03$ (C500), confirming that non-extensivity is intrinsic to the system.

\section{\label{sec:Theoretical-framework}THEORETICAL FRAMEWORK AND PHYSICAL INTERPRETATION}
In this section, we focus on the probability distribution models used to fit the data, emphasizing their similarities and the applied physical constraints. Table \ref{tab:nomenclature_main} summarizes the nomenclature and key equations. Discrete versions are presented in the Methods section, with further details provided in the supplementary material.

We pay special attention to the $q$-Gaussian. To our knowledge, the Tsallis $q$-Gaussian has not been explicitly derived in a 3D radial form nor applied to fit the end-to-end distance distributions of polymer chains in melts generated by atomistic molecular dynamics simulations. Since the original works are written in a somewhat abstract mathematical formalism, we provide a brief description of the theory in Appendix \ref{app:qgauss-derivations} and supplementary material (Sec.~S1), highlighting the details relevant to understanding its application to polymer end-to-end statistics. We also emphasize the limitations of this distribution function when modeling polymer systems. The final section addresses entropy evaluation, comparing the classical BG entropy with the Tsallis $q$-entropy.

\subsection{\label{subsec:theory-gaussian}Radial Gaussian probability density in 3D}
This is the classical model for an ideal chain of freely jointed Gaussian $N_g$ segments (where $N_g = N_k$)\footnote{The minimal statistical segment is the Kuhn segment. The subscript $g$ is preferred here to differentiate the two distributions, Gaussian and $q$-Gaussian} of length $l_g (= l_k)$.\cite{Teraoka-PolymSol-2002,Rubinstein-Colby-PolPhys-2003}
Starting from the volumetric probability density $\rho_g(\mathbf{R})$ (units of $L^{-3}$), the radial probability density $P_g(R)$ for the distance $R =|\mathbf{R}|$ (units of $L^{-1}$) is defined such that $P_g(R)dR$ is the probability of finding the chain end-to-end distance in the interval $[R, R+dR]$. It is related to the volumetric probability density by $P_g(R) = 4\pi R^2 \rho_g(\mathbf{R})$, leading to:
\begin{align}
	P_{\text{g}}(R;N_\text{g},l_\text{g}) & = 4\pi R^2\left(\frac{3}{2\pi N_g l_g^2}\right)^{3/2}	\exp\!\left[-\frac{3R^2}{2N_g l_g^2}\right] \nonumber\\
	& = C_g R^2 \exp\!\left(-\beta_g R^2\right)\, ,
	\label{eq:probability-Gauss_radial}
\end{align}
with the normalization constant $C_g = 4 \pi \left(\beta_g/\pi\right)^{3/2}$ and the scaling parameter $\beta_g = 3/(2N_g l_g^2)$ that determine the width of the distribution.

The continuous distribution is normalized over all $R$, $\int_{0}^{\infty} P_g(R)\, dR = 1$. The second moment is:
\begin{equation}
	\langle R^2\rangle_g = \int_{0}^{\infty} R^2 P_g(R) dR = N_g l_g^2 = \frac{3}{2\beta_g}\, .
	\label{eq:R^2-gaussian}
\end{equation}
Another important result, used in the evaluation of the entropy of a Gaussian chain, is the linearity in $R^2$ of the logarithm of
the volumetric probability density $\rho_g(\mathbf{R}) \propto P_g(R)/R^2$,
\begin{equation}
	\ln\; \left[\frac{P_{\text{g}}(R)}{R^2}\right] = \ln(C_g) - \beta_g R^2 \, ,
	\label{eq:propoto-R^2}
\end{equation}
Both $P_{\text{g}}(R)/R^2$ and the normalization constant $C_g$ have units of $L^{-3}$. The linearity in $R^2$ expressed by Eq.~(\ref{eq:propoto-R^2}) allows for a straightforward assessment of Gaussianity. In this expression, as in subsequent equations, the ratio $P_{\text{g}}(R)/R^2$ is implicitly normalized by a reference density $\rho_0 = 1$ (also with units of $L^{-3}$) to ensure the dimensionless nature of the logarithm argument.

\subsection{\label{subsec:theory-q-gaussian}The $q$-Gaussian and its properties}
Within the framework of nonextensive statistical mechanics (NSM), Tsallis\cite{Tsallis-1998, Gell-Mann-Tsallis-2003, Tsallis-2009} introduced two generalized functions, the ``$q$-exponential'' and the ``$q$-logarithm'', together with the concept of an escort distribution used to compute finite moments when $q\ne 1$ and the standard methods fail. In this work we do not use escort distributions, since for the range of $q$ values of interest ($0 < q < 5/3$ in three dimensions) the moments we need remain finite. A brief summary of the $q$-mathematics was provided by \citeauthor{Umarov-2008}.\cite{Umarov-2008}

We define below the $q$-exponential function and the normalized radial $q$-Gaussian distribution, emphasizing their main properties and physical implications for polymer end-to-end distance statistics. Both the $q$-exponential and $q$-logarithm functions reduce to the corresponding exponential and logarithm functions when $q=1$. We focus below on the key equations transferring derivation details to Appendix \ref{app:qgauss-derivations} and the supplementary material, Sec.~1.

\subsubsection{\label{subsubsec:theory-q-exp}The $q$-exponential function}
In nonextensive statistical mechanics Tsallis introduced a generalized exponential function, the $q$-exponential [$\exp_q(u)$] as a generalization of [$\exp(u)$].\cite{Tsallis-1998, Gell-Mann-Tsallis-2003, Tsallis-2009} It is defined for a single argument $u$ as
\begin{equation}
	\exp_q(u)  = [\,1 + (1 - q)u\,]_+^{\frac{1}{1-q}} \, , 
	\label{eq:q-exp}
\end{equation}
where the operator $[\cdot]_+$ ensures that the argument remains positive, guaranteeing real values of the $q$-exponential.
\footnote{The Tsallis derivation is straightforward, as demonstrated on page 5 of Ref.~\onlinecite{Gell-Mann-Tsallis-2003}.
While the solution to the linear differential equation $dy/dx = y$ with the initial condition $y(0)=1$ is the standard exponential $y = e^x$, the solution to the non-linear case $dy/dx = y^q$ (where $y(0)=1$ 	and $q \in \mathbb{R}$) defines the $q$-exponential function $y = e_q^x$.
Correspondingly, its inverse is defined as the 	$q$-logarithm, $y = \ln_q(x)$.
If $y$ is some quantity,  a population or a probability, ..., and $x$, the independent variable, time, end-to-end distance, ..., the linear equation $dy/dx = y$ indicates that growth rate of the population depends on the population size $y$, with no memory of the path taken to reach it. This means statistical independence and consequentially extensivity.
In the non-linear case, the rate of change depends non-linearly on the current state, $(dy/y)/dx = y^{q-1}$: the increment $dy$ produced by $dx$ is amplified or suppressed relative to the linear case depending on whether $q>1$ or $q<1$. Parts os the system interact so strongly that the whole is not just the sum of its parts. Statistical independence is lost and entropy is non-extensive.   
}
This constraint defines the domain of the function for $q<1$. We will drop this subscript in the equations below.

In the $q$-Gaussian distribution $P_q(R)$, the argument is $u = - \beta_q R^2$, where $R$ is the end-to-end distance of polymer chains and $\beta_q$ plays the role of an inverse variance, being formally similar to $\beta_g$ defined in Sec.~ \ref{subsec:theory-gaussian} (see also Table \ref{tab:nomenclature_main}). The general form of the $q$-exponential for the $q$-Gaussian is then 
\begin{equation}
	\exp_q(- \beta_q R^2)  = [\,1 - (1 - q)\beta_q R^2\,]^{\frac{1}{1-q}}\, . 
	\label{eq:q-exp-for-q-Gaussian}
\end{equation}

There are three situations to consider: (i) $q = 1$ where the $q$-Gaussian reduces to the Gaussian and the entropy is extensive, (ii) $q < 1$ where Eq. (\ref{eq:q-exp-for-q-Gaussian}) is well defined within finite bounds and the distribution has compact support (supplementary material, Sec.~ S1.1), and (iii) $q > 1$ where the distribution has infinite support and long tails (see Appendix \ref{app:qgauss-derivations} and supplementary material, Sec.~ S1.2). The entropy is non-extensive for both $q<1$ and $q>1$, being superadditive for the former ($S_{system}>\sum S_{parts}$) and subadditive for the latter. 

For $q < 1$, since $(1 - q)>0$, the expression inside brackets in Eq. (\ref{eq:q-exp-for-q-Gaussian}) reaches zero at a finite distance
\begin{equation}
	R_\text{max} =\sqrt{1/\beta_q(1-q)},
	\label{eq:r_max-compact-support}
\end{equation}
giving the distribution compact support [$P_q(R) = 0$ for $R > R_\text{max}$].

For $q > 1$, the same expression, Eq. (\ref{eq:q-exp-for-q-Gaussian}), can be rewritten as a decaying power law by noting that $1-q < 0$. Defining $q^\prime = q > 1$, the exponent becomes negative, $1/(1-q^\prime) = - 1/(q^\prime - 1)$, yielding a long-tailed distribution. The result is a decaying function in $R$,
\begin{equation}
	\exp_q(- \beta_q R^2)  = [\,1 + (q - 1)\beta_q R^2\,]^{-\frac{1}{q-1}}\,. 
	\label{eq:q-exp-for-q-Gaussian-q>1}
\end{equation}

Notice the different forms of Eqs. (\ref{eq:q-exp-for-q-Gaussian}) and (\ref{eq:q-exp-for-q-Gaussian-q>1}). They are two equivalent parametrizations of the same family of functions, convenient for discussing separately the compact-support ($q<1$) and long-tailed ($q>1$) regimes. Both representations are mathematically equivalent and describe a single universal family of distributions, the $q$-Gaussian, which reduces to the classical Gaussian in the limit $q\, \to\, 1$.

\subsubsection{\label{subsubsec:theory-radial-q-gaussian}Radial $q$-Gaussian probability density in 3D}
The radial $q$-Gaussian probability density (with units of $L^{-1}$) for the scalar end-to-end distance $R$ is defined by
\begin{equation}
	P_q(R) = C_q R^2 \exp_q\left(-\beta_q R^2\right),
	\label{eq:q-Gaussian-general form-radial}
\end{equation}
where $C_q$ is the normalization constant (units $L^{-3}$) ensuring $\int_0^\infty P_q(R)\, dR = 1$. For $q<1$ the distribution has compact support, while for $q>1$ it exhibits long tails (Appendix \ref{app:qgauss-derivations} and supplementary material, Sec.~1.2).

\subsubsection{Second moment}
\label{subsubsec:theory-second-moment}
Since the moments of interest remain finite for the range of $q$ values relevant to the polymer melts studied in this work ($q<5/3$ in three dimensions), we employ the standard expectation value
\begin{equation}
	\langle R^k \rangle_q = \int_0^{\infty} R^k P_q(R)\, dR,
	\label{eq:moment definition}
\end{equation}
avoiding the use of escort distributions. 

As shown in the Appendix (Sec.~\ref{app:qgauss-second-moment}) and supplementary material (Sec. S1.2), for both cases ($q<1$ and $q>1$), the mean-square end-to-end distance of a $q$-Gaussian random walk is
\begin{equation}
	\langle R^2 \rangle_q  = \frac{3}{\beta_q \left(7-5q\right)}\; .
	\label{eq:R^2-q-gaussian-final}
\end{equation}

\subsubsection{Mapping of the q-Gaussian second moment onto an equivalent random walk}
\label{subsubsec:theory-mapping}
To relate this $q$-Gaussian mean-square end-to-end distance to an effective random walk, we define
\begin{equation}
	\beta_q = \frac{3}{7-5q} \, \frac{1}{N_q\, l_q^2}, 
	\label{eq:beta_q-definition}
\end{equation}
that equals $\beta_g$ when $q = 1$. Inserting this definition in Eq. (\ref{eq:R^2-q-gaussian-final}), the second moment may be expressed as a random walk of $N_q$ steps, each with length $l_q$
\begin{equation}
	\langle R^2 \rangle_q  = N_q l_q^2.
	\label{eq:q-generalized-rd-walk}
\end{equation}

It is important to clarify the physical meaning of this construction. With the definition of $\beta_q$ in Eq.~(\ref{eq:beta_q-definition}), the second moment $\langle R^2\rangle_q$ is constrained to the same random-walk form as in the Gaussian case expressed by Eq.~(\ref{eq:R^2-gaussian}). 
However, this does not imply that the microscopic heterogeneity of segment types (ACS/RCS) is absent or irrelevant. 
The non-Gaussianity and non-extensivity are manifested in $P_q(R)$ through the entropic index $q$ in the normalization constant and the $q$-exponential function. 

\subsubsection{The $q$-logarithm and linearization}
\label{subsubsec:theory-q-log}
In the Gaussian case ($q=1$), as shown in Eq.~(\ref{eq:propoto-R^2}), the plot of the logarithm of the volumetric density $\rho_g(R) \propto P_g(R)/R^2$ versus $R^2$ is linear. For the $q$-Gaussian, this linearity is preserved only when utilizing the $q$-logarithm.
The $q$-logarithm is the inverse function of the $q$-exponential [$\ln_q[\exp_q (u)] = u$] and is defined for $y>0$ as
\begin{equation}
	\ln_q(y)= \frac{y^{1-q}-1}{1-q}\, .
	\label{eq:q-logarithm}
\end{equation}
The $q$-logarithm  satisfies a \textit{pseudo-additive} property for the product of two variables $x$ and $y$: $\ln_q(xy) = \ln_q(x) + \ln_q(y) + (1-q)\ln_q(x) \ln_q(y)$. \cite{Gell-Mann-Tsallis-2003}

For the $q$-Gaussian, linearity in $R^2$ is preserved when using the $q$-logarithm. Applying Eq.~(\ref{eq:q-logarithm}) to Eq.~(\ref{eq:q-Gaussian-general form-radial}) yields
\begin{equation}
	\ln_q\left[\frac{P_q(R)}{C_q R^2}\right] = -\beta_q R^2 \, .
	\label{eq:propoto-R2-q-gaussian-clean}
\end{equation}
To facilitate the comparison  with the Gaussian case  this equation may also be expressed as
\begin{equation}
	\ln_q\left[\frac{P_q(R)}{R^2}\right] = \ln_q(C_q) - \beta_q R^2 \left[1 + (1-q)\ln_q(C_q)\right] \, .
	\label{eq:propoto-R2-q-gaussian}
\end{equation}
For this derivation we use Eq. (\ref{eq:propoto-R2-q-gaussian-clean}) and apply the  pseudo-additive property of $\ln_q$ to the product $C_q \cdot \left(P_q(R)/(C_q R^2)\right)$.
While $\ln_q[P_q(R)/R^2]$ remains linear in $R^2$, the slope is ``renormalized'' by the factor $[1 + (1-q)\ln_q(C_q)]$. With Eq. (\ref{eq:propoto-R2-q-gaussian}) we may evaluate the entropic index $q$ and the scaling parameter $\beta_q$ directly from simulation data.

\subsection{Entropy formalisms: Boltzmann--Gibbs and Tsallis}
\label{subsec:theory-entropy}

We calculate entropy directly from the normalized radial probability density $P(R)$ to quantify the number of accessible chain conformations, avoiding the use of escort distributions. 

\begin{table*}[t]
	\centering
	\caption{\label{tab:nomenclature_main}Nomenclature and analytical expressions for Gaussian and $q$-Gaussian models. The discrete equations for the probability and entropy were used in the numerical evaluations. }
	\renewcommand{\arraystretch}{1.8} 
	\begin{ruledtabular}
		\begin{tabular}{lll}
			\textbf{Quantity} & \textbf{Gaussian ($q=1$)} & \textbf{$q$-Gaussian (general $q$)} \\
			\hline
			Volumetric prob. density & \multicolumn{2}{l}{$\rho(R)$ [Units: \AA$^{-3}$]} \\
			Radial prob. density     & \multicolumn{2}{l}{$P(R) = 4\pi R^2 \rho(R)$ [Units: \AA$^{-1}$]} \\
			Discrete probability     & \multicolumn{2}{l}{$p_k = P(R_k) \Delta R$ [dimensionless]} \\
			\hline
			Radial probability density & 
			$P_{\mathrm{g}}(R) = C_g R^2 \exp(-\beta_g R^2)$ &
			$P_q(R) = C_q R^2 \exp_q(-\beta_q R^2)$ \\
			& & $\quad \quad \; \; = C_q R^2 [1 - (1-q)\beta_q R^2]^{1/(1-q)}$ \\
			\hline
			Normalization constant & 
			$\displaystyle C_g = 4\pi \left(\frac{\beta_g}{\pi}\right)^{3/2}$ &
			$\displaystyle C_q = 4\pi \left(\frac{\beta_q}{\pi}\right)^{3/2} \times 
			\begin{cases} 
				(1-q)^{3/2} \frac{\Gamma\left(\frac{5}{2}+\frac{1}{1-q}\right)}{\Gamma\left(1+\frac{1}{1-q}\right)}, & q < 1 \\[10pt]
				(q-1)^{3/2} \frac{\Gamma\left(\frac{1}{q-1}\right)}{\Gamma\left(\frac{1}{q-1}-\frac{3}{2}\right)}, & q > 1 
			\end{cases}$ \\ [10pt]
			\hline 
			Support (range in $R$) & 
			$R \in [0, \infty)$ &
			$\begin{cases}
				R \in [0, R_{\max}], & q<1, \, R_{\max}=\sqrt{\frac{1}{\beta_q(1-q)}}\\[6pt]
				R \in [0, \infty), & q>1
			\end{cases}$ \\ [10pt]
			\hline
			Logarithmic form & 
			$\displaystyle \ln\left[\frac{P_g(R)}{R^2 \rho_0}\right] = \ln(C_g) - \beta_g R^2$ &
			$\displaystyle \ln_q\left[\frac{P_q(R)}{R^2 \rho_0}\right] = \ln_q(C_q) - \beta_q R^2[1+(1-q)\ln_q(C_q)]$ \\
			\hline
			Second moment & 
			$\displaystyle \langle R^2\rangle_g = \frac{3}{2\beta_g}$ &
			$\displaystyle \langle R^2\rangle_q = \frac{3}{(7-5q)\beta_q}$ \\
			\hline
			Effective inverse variance & 
			$\displaystyle \beta_g = \frac{3}{2N_g l_g^2}$ &
			$\displaystyle \beta_q = \frac{3}{(7-5q) N_q l_q^2} $ \\
			\hline
			Entropy (continuous) & 
			$\displaystyle S_1 = -k_B \int_0^\infty P(R) \ln\left[\frac{P(R)}{4\pi R^2 \rho_0}\right] dR$ &
			$\displaystyle S_q = \frac{k_B}{q-1} \left( 1 - \int_0^\infty P(R) \left[ \frac{P(R)}{4\pi R^2 \rho_0} \right]^{q-1} dR \right)$ \\
			\hline
			Entropy (discrete) & 
			$\displaystyle S_1 = -k_B \sum_k p_k \ln p_k$ &
			$\displaystyle S_q = k_B \frac{1 - \sum_k p_k^q}{q-1}$ \\
			\hline
			Entropy character & 
			Extensive: $S(A+B)=S(A)+S(B)$ &
			Pseudo-additive: $\displaystyle S_q(A+B)=S_q(A)+S_q(B)+\frac{1-q}{k_B}S_q(A)S_q(B)$ \\
		\end{tabular}
	\end{ruledtabular}
\end{table*}

\subsubsection{\label{subsubsec:B-G-entropy}Boltzmann--Gibbs entropy ($S_1$)}
The classical extensive entropy is defined using the volumetric probability density $\rho(R)$ (units of $L^{-3}$) integrated over the 3D volume. Expressed in terms of the radial probability density $P(R)$ (units of $L^{-1}$), it is:
\begin{equation}
	S_1 = -k_\text{B} \int_{0}^{\infty} P(R) \ln \left[ \frac{\rho(\mathbf{R})}{\rho_0} \right] dR + \text{const.} \, ,
	\label{eq:S1_def}
\end{equation}
where $\rho(\mathbf{R}) = P(R)/(4\pi R^2)$ and $\rho_0 = 1$~\text{\AA}$^{-3}$ is a reference density introduced to ensure a dimensionless argument for the logarithm. 

For applying Eq. (\ref{eq:S1_def}) to a normalized histogram for the end-to-end distances, we use the Gibbs probabilistic definition of entropy. 
The dimensionless probability for each bin is 
$p_k = \rho(R_k) dV(R_k) = P(R_k) \Delta R$,
which corresponds to the probability of finding the end-to-end distance within the spherical shell of volume $V_k \approx 4\pi R_k^2 \Delta R$. The discrete entropy is 
\begin{equation}
	S_1 = -k_\text{B} \sum_{k} p_k \ln p_k \, .
	\label{eq:S1_discrete}
\end{equation}
Extensivity and additivity emerge when statistically independent subsystems are considered. As discussed further in the Methods section, the discrete version, Eq. (\ref{eq:S1_discrete}), differ from the continuous version, Eq. (\ref{eq:S1_def}), by a constant offset, $ \ln(\Delta R)$, that is canceled out in the non-extensivity evaluation made in this work. 

\subsubsection{\label{subsubsec:Tsallis-entropy}Tsallis $q$-entropy ($S_q$)}
To capture possible non-extensive behavior due the existence of different types and relative populations  of Kuhn segments, we use the generalized Tsallis entropy.\cite{Tsallis-2009,Lyra-Tsallis-Gell-Mann-2004} In the discrete implementation, the dimensionless probability $p_k$ is raised to the power $q$ 
\begin{equation}
	S_q = k_\text{B} \frac{1 - \sum_k p_k^q}{q - 1} \,.
	\label{eq:Sq_discrete}
\end{equation}
As $q \to 1$, $S_q$ converges to $S_1$. A value of $q \neq 1$ signals a departure from the extensive, BG framework. As with the Gaussian, replacing in Eq. (\ref{eq:Sq_discrete}) the discrete probability for each bin $p_k$, we have the equation applied to the normalized histogram of end-to-end distances
\begin{equation}
	S_q = k_\text{B} \frac{1 - \sum_k \left[P(R_k) \Delta R \right]^q   }{q - 1} \, .
	\label{eq:Sq_discrete-pk-bin}
\end{equation}

For statistically independent subsystems $A$ and $B$, with $p_{ij}^{A+B} = p_i^A p_j^B$, the Tsallis entropy obeys a non-additive composition rule
\begin{equation}
	S_q(A+B) = S_q(A) + S_q(B)
	+ \frac{1-q}{k_\text{B}}\,S_q(A)\,S_q(B).
\end{equation}
The additional term is proportional to $(1-q)$ and contains a factor $1/k_\text{B}$ for dimensional consistency, since each $S_q$ has units of 	$k_\text{B}$. 
Therefore, for $q<1$ the entropy is superadditive ($S_q(A+B) > S_q(A)+S_q(B)$), while for $q>1$ it is subadditive ($S_q(A+B) < S_q(A)+S_q(B)$) in this idealized case of independent subsystems. 

In this work, the entropic index $q $ has the role of an \textit{heterogeneity-index}.  It quantifies the deviation of the end-to-end distance statistics from the extensive Boltzmann–Gibbs (BG) limit $q=1$, providing also information on the relative prevalence of different Kuhn-segment types (ACS vs. RCS) and their impact on the chain's global statistical and thermodynamic properties.

\section{Methods}
\label{sec:methods}
\subsection{\label{subsec:methods:MD sim} Molecular dynamics simulation details}

The molecular dynamics simulations were performed using the methodology detailed in previous works. \cite{Martins-macromol-2013, Martins-2026} All systems were modeled in GROMACS\cite{Hess-2008a, Hess-1997a} using the united-atom force field of Paul \textit{et al.}\cite{Paul-1995a}, with periodic boundary conditions in the $NpT$ ensemble at 600 K and 1 bar.  The simulation box is cubic (side $\approx 160$ \AA), containing 140000 united atoms of \ce{CH2}, covering unentangled (C50 and C100) and entangled (C250 and C500) chains. The molar mass of the polyethylene chains varies from 700 \si{g.mol^{-1}} for C50 up to 7000 \si{g.mol^{-1}} for C500.

Literature values for the entanglement molar mass of polyethylene,  $M_e$, typically range from 800 to 1200 \si{g.mol^{-1}}. \cite{Liu-Ruymbeke-2006, Fetters-2007} Consequently, the C50 chains are unequivocally unentangled, the C250 contain between three to four entanglements per chain, while the C500 chains contain around six to eight entanglements per chain (depending on the $M_e$ values considered). 

All systems were equilibrated and the production runs extend beyond the chain end-to-end orientational autocorrelation time (e.g. $\approx3$ ns for C250), ensuring reliable statistics for the calculation of end-to-end distance distributions. Frames were saved every 1 ps (0--1 ns), every 10 ps (1--30 ns), and every 100 ps thereafter, yielding $\sim$ 4600 frames per trajectory. 

\subsection{\label{subsec:methods:data-histograms}End-to-end distance histograms}

For each saved frame $t$ and chain $i$ ($i = 1, \ldots, N_\text{ch}$), we computed the unwrapped end-to-end distance $R_{i,t}=|\mathbf{R}_{i,t}|$. These distances were discretized into radial bins of width $\Delta R=1$~\AA. The bin centers $R_k$ ($k = 0, \cdots, N_{\text{bin}-1}$) cover the interval $[0, R_\text{max}]$, where $R_\text{max}$ is the largest distance observed in the trajectory.

The histograms $H_t(R_k)$ (per frame) or $H(R_k)$ (aggregate) are converted to the normalized radial probability density $P(R_k)$ (units of \AA$^{-1}$) by
\begin{subequations}
	\begin{align}
		\rho(R_k)&= \frac{H(R_k)}{\sum_j H(R_j)} \,  \frac{1}{4\pi R_k^2\,\Delta R}\,
		\label{eq:discrete:volume-prob-desity-rho}
		\\
		P(R_k) &= \frac{p_k}{\Delta R}
		\label{eq:discrete:radial-prob-density}
	\end{align}
\end{subequations}
where the term $p_k = H(R_k)/\sum H(R_j)$ represents the dimensionless discrete probability of finding a chain in bin $k$.  By construction, $P(R_k) = 4\pi R_k^2 \rho(R_k)$. Consequently,
$p_k = \rho(R_k)\,4\pi R_k^2 \Delta R = P(R_k)\Delta R $,
which satisfies the normalization $\sum_k p_k = 1$. All statistical moments are computed using these discrete probabilities as
\begin{equation}
	\langle R^n \rangle = \sum_k R_k^n\,p_k\, .
\end{equation}

This dimensional consistency in the probability equations [(\ref{eq:discrete:volume-prob-desity-rho}) and (\ref{eq:discrete:radial-prob-density})] and in the the dimensionless probability $p_k$, emphasized also in Table \ref{tab:nomenclature_main}, 
is particularly critical for numerical evaluations. In the processing of simulation data, the conversion of discrete histogram counts $H(R_k)$ into the radial density $P(R_k)$ via Eq.~(\ref{eq:discrete:radial-prob-density}) ensures that the radial probability density is independent of the specific bin width $\Delta R$. 
Furthermore, the explicit identification of the volumetric probability density $\rho(R) = P(R)/(4\pi R^2)$ allows for a  numerical fit of Eqs.~(\ref{eq:propoto-R^2}) and (\ref{eq:propoto-R2-q-gaussian}) to the data less sensitive to the large dynamic range of the probability values near the distribution's tail. 

Further details on the data handling, the construction of the data histograms and evaluation of the ``experimental'' moments are provided in the supplementary material, Sec.~ S2.1. A snippet of the code used is also provided as Code S2. 

\subsection{\label{subsec:methods:Sp-and-Ti-Fr-by-Fr-final}Ensemble constructions}

We considered two averaging approaches: a global averaging of all trajectory files (Space and Time, Sp\&Ti) and a per-frame averaging (Frame-by-Frame, FrbyFr). Both approaches yield the same distribution function for the end-to-end distance, and the resulting fits coincide within statistical uncertainty.

\subsubsection{\label{subsubsec:Space-and-Time} Space and time (Sp\&Ti) averaging}
In this conventional approach (typical of steady-state molecular dynamics sampling), all sampled chain configurations in space and time are combined into a single ensemble. Every chain $i$ ($i = 1, \ldots, N_\text{ch}$), at every frame $t$, contributes an end-to-end distance $R_{i,t}$ to the overall distribution. The space-time probability distribution is the normalized histogram of all these distances:
\begin{equation}
	P_{\text{Sp\&Ti}}(R) = \frac{1}{N_\text{ch} N_\text{t}} \sum_{i=1}^{N_\text{ch}} \sum_{t=1}^{N_\text{t}} \delta(R - R_{i,t}) \, .
	\label{eq:probability-Sp-and-Ti}
\end{equation}
This represents an ergodic ensemble average, where $\delta$ is the Dirac delta function implemented via histogram binning, and $\int P_{\text{Sp\&Ti}}(R) \, dR = 1$. The mean-square end-to-end distance is $\langle R^2\rangle_{\mathrm{Sp\&Ti}} = \int R^2 P_{\mathrm{Sp\&Ti}}(R)\,dR$. For an illustration of this procedure see the Code snippet S2 in the supplementary material. The key result is the generation of a grid of distance values (\texttt{R\_fit}$\equiv R_k$ )  and the corresponding probabilities (\texttt{P\_fit}$\equiv P_k$). 

\subsubsection{\label{subsubsec:frame-by-frame}Frame--by--frame (FrbyFr) averaging}
In this approach, each simulation frame is treated as an independent, quasi-equilibrium instantaneous configuration. The molecular dynamics thermostat maintains thermal equilibrium even as conformational variables fluctuate.

In each frame there are fast and slow events. The fast events are the changes on local chain conformations with characteristic timescales such as  the orientational relaxation time of the different types of Kuhn segments ($\tau_\text{k,ACS}, \tau_\text{k,RCS}$ and $\tau_\text{k,CE}$) quantified in a previous work, where $\tau_\text{k,ACS}\approx10\tau_\text{k,RCS}$.\cite{Martins-2026} Much slower events are, for example, the local alignment of chains segments participating in ordered regions in the melt, or the fluctuations in $\langle R^2 \rangle$.\footnote{Specifically, this last fluctuation is quantified with the orientational end-to-end chain autocorrelation time  ($\tau_\text{ch}$),  obtained fitting the $\langle \mathbf{R}(t) \cdot \mathbf{R}(0) \rangle / \langle R^2(0) \rangle$ to a relaxation function, typically a stretched exponential.} 

The sampling interval must resolve these slow fluctuations, while the total simulation time should span multiple autocorrelation cycles to ensure good statistics. For frame $t$, the distribution of end-to-end distances across chains (space sampling) is
\begin{equation}
	P_t(R) = \frac{1}{N_\text{ch}} \sum_{i=1}^{N_\text{ch}} \delta(R - R_{i,t}),
	\label{eq:probability_t-Fr-by-Fr}
\end{equation}
with its own instantaneous mean-square distance $\langle R^2 \rangle_t = \int R^2 P_t(R)\,dR$.

The overall FrbyFr distribution is the average of these instantaneous distributions:
\begin{equation}
	P_{\text{FrbyFr}}(R) = \frac{1}{N_\text{t}} \sum_{t=1}^{N_\text{t}} P_t(R),
	\label{eq:probability-P(R)-Fr-by-Fr}
\end{equation}
representing an average over fluctuating local equilibrium states, where $\int P_{\text{FrbyFr}}(R) \, dR = 1$. This is mathematically identical to $P_{\text{Sp\&Ti}}(R)$ only if $\langle R^2\rangle_t$ remains constant over time.

\subsection{\label{subsec:methods:model-fits}Model fits: radial Gaussian and
	$q$-Gaussian}

To critically compare the Gaussian and q-Gaussian models, we must account for their different degrees of freedom and their sensitivity to tail deviations.
The q-Gaussian includes one additional parameter ($q$) to capture the non-extensivity
arising from the mosaic of ACS and RCS segments. 
As previously shown, the Gaussian model fails for unentangled chains\cite{Martins-2026}
and for entangled chains in the large-$R$ limit, where extended ACS configurations prevail (Sec. \ref{subsec:results:ETE-distribution}).

Model implementations strictly follow the derivations in Sec. \ref{sec:Theoretical-framework} and Appendix \ref{app:qgauss-derivations}. Specific procedures for the Gaussian (Codes S3–S4) and q-Gaussian (Codes S5–S7) are detailed in the supplementary material (Sec. S2.2).

We evaluate model quality using four complementary criteria that probe distinct aspects of the distributions:
(i)~the reduced chi-squared statistic $\chi^{2}/\nu$,\cite{Bevington-2003,Press-2007} where $\nu = n - m$ is the number of degrees of freedom for a fit of $n$ data points (histogram bins) with $m$ free parameters ($m = 1$ for the Gaussian and $m = 2$ for the $q$-Gaussian);
(ii)~the root-mean-square error (RMSE) of the histogram residuals;
(iii)~the non-Gaussianity parameter $\alpha_{2}$; and 
(iv)~the root-mean-square error of the quantile--quantile (Q--Q) plot, $\mathrm{RMSE}_{\mathrm{QQ}}$.
Additional information on these criteria is provided in Sec. S4 (supplementary material). 

Given the large sample sizes used in this work ($N \gtrsim 10^{6}$ chain configurations), even small systematic deviations result in large absolute $\chi^{2}$ values.
Consequently, $\chi^{2}/\nu$ is treated as a relative value for model comparison rather than as an absolute quantifier of fit quality. 
The parameters $\alpha_{2}$, the RMSE and  Q--Q plots were described and used previously to classify non-Gaussianity and transition to Gaussianity in Kuhn segment sequences.\cite{Martins-2026}
Table \ref{tab:fit quality criteria} summarizes the sensitivity and physical information provided by each criterion. 
Their convergence toward a single conclusion provides evidence for the preference of one model over the other.\cite{Bevington-2003,Thode-2002}

\begin{table*}[htbp]
	\centering
	\caption{%
		Summary of the diagnostic tests used to compare the Gaussian (\textbf{G}) and $q$-Gaussian (\textbf{$q$-G}) fits to the data providing complementary information. 
		For further details see the supplementary material, Sec. S4. 
	}
	\label{tab:fit quality criteria}
	\renewcommand{\arraystretch}{1.35}
	\begin{ruledtabular}
	\begin{tabular}{llll}
		Group & Diagnostic & Sensitivity & Gaussian failure signature \\
		\hline 
		Global
		& $\chi^{2}/\nu$
		& Overall histogram fit
		& $\chi^{2}/\nu \gg 1$ \\[2pt]
		& RMSE
		& Residual magnitude
		& Large absolute residuals \\
		\midrule
		Moments
		& $\alpha_{2}$
		& Compact support
		& $\alpha_{2}^{\mathrm{G}} = 0$; data $< 0$ \\[2pt]
		& Skewness, kurtosis
		& Distribution shape
		& Asymmetry, non-zero kurtosis \\
		\midrule
		Q--Q shape
		& $\mathrm{RMSE}_{\mathrm{QQ}}$
		& Tail quantile errors
		& Curvature at large $R$ \\[2pt]
	\end{tabular}
	\end{ruledtabular}
\end{table*}

\subsection{\label{subsec:methods:tail}Tail definition and tail concentration}

To analyze the contribution of extended conformations to the mean-square end-to-end distance, we define the ``tail'' of the distribution as the region exceeding the 75th percentile
($Q_{75}$)  of the per-frame radial probability density P(R).
For each simulation frame $t$, we evaluate the empirical cumulative distribution function $Ft(Rk)$ from the discrete probabilities $p_k^{(t)} = P_t(R_k)\,\Delta R$:
\begin{equation}
	F_t(R_k) = \sum_{j:\, R_j \le R_k} p_j^{(t)}\; .
\end{equation}
The limiting $R_{Q_{75}}^{(t)}$ is obtained via linear interpolation between neighboring bins at $F_t(R) = 0.75$.
The tail concentration, $T_2^{(t)}$, is then defined as the fraction of $\langle R^2 \rangle$ arising from chains in the last quartile:
\begin{equation}
	T_2^{(t)} = 
	\frac{\displaystyle\int_{R_{Q_{75}}^{(t)}}^{\infty} R^2\,P_t(R)\,dR}
	{\displaystyle\int_0^{\infty} R^2\,P_t(R)\,dR}\,.
	\label{eq:T2-definition}
\end{equation}
Physically, the $R^2$ weighting in Eq. (\ref{eq:T2-definition})  re-weights the distribution toward extended configurations. 
While the fraction of chains in the tail ($f_{\text{tail}}= \int_{\mathrm{Q_75}}^{\infty}P(R) dR = 0.25$ ) represents only 25\% of the chains, they provide the dominant contribution for the second moment of the distribution. 
For a theoretical Gaussian distribution, this calculation yields $T_2 = 53.4\% $ (Sec. \ref{subsec:results:tail-concentration}).

To ensure consistency, $T_2^{(t)}$ for the Gaussian and $q$-Gaussian models is evaluated on the same grid $R_k$ used for the histograms (Sec.~\ref{subsec:methods:Sp-and-Ti-Fr-by-Fr-final}).
This allows a direct, frame-by-frame comparison between the tail concentration obtained from the simulation data and that predicted by each model distribution.

\subsection{\label{sec:methods:entropy-quantification}Entropy and non-extensivity quantification}

The entropy values are computed directly from the per-frame discrete probabilities $p_k^{(t)} = P_t(R_k)\,\Delta R =  \rho_t(R_k) 4 \pi R_k^2 \Delta R$, where $P_t(R_k)$ is the normalized radial probability density of frame $t$ and  $\sum_k p_k^{(t)} = 1$. 
For each frame $t$, we evaluate the Boltzmann--Gibbs entropy $S_1^{(t)}$ and the Tsallis $q$-entropy $S_q^{(t)}$ using the discrete forms of Eqs. (\ref{eq:S1_discrete}) and (\ref{eq:Sq_discrete-pk-bin}):
\begin{equation}
	S_1^{(t)} = -k_B \sum_k p_k^{(t)} \ln p_k^{(t)}\,,
	\label{eq:S1_frame}
\end{equation}
\begin{equation}
	S_q^{(t)} = k_B\,\frac{1 - \sum_k \bigl[p_k^{(t)}\bigr]^{q}}{q - 1}\,.
	\label{eq:Sq_frame}
\end{equation}
We emphasize that the index $q$ in Eq. \eqref{eq:Sq_frame} is the frame-specific $q_{\text{fit}}^{(t)}$ obtained from the $q$-Gaussian fit to  $P_t(R)$. 
Thus, both the probability set $ \left\{p_k\right\}   $ and the entropic index $q$ are updated per frame. 
The volume probability density is evaluated in Code S8 and the numerical implementation of Eqs. \eqref{eq:S1_frame} and \eqref{eq:Sq_frame} is described in Code S9 (supplementary material, Sec. S2.3). 

The transition from continuous to discrete entropy introduces bin-width-dependent offsets (supplementary material, Sec.~S3):
\begin{align}
	S_{1,\mathrm{cont}} &= S_{1,\mathrm{disc}} + k_B \ln(\Delta R),
	\label{eq:S1_offset} \\
	S_{q,\text{cont}} &= (\Delta R)^{1-q} S_{q,\text{disc}} + k_\text{B} \left[ \frac{1 - (\Delta R)^{1-q}}{q-1} \right].
	\label{eq:Sq_offset}
\end{align}
By fixing the bin width at exactly $\Delta R = 1$\,\AA, we establish a reference state with $\rho_0 = 1$\,\AA$^{-3}$.
Under this condition $\ln(\Delta R) = 0$ and $\left(\Delta R \right)^{1-q} = 1$, causing the additive offset in Eq. \eqref{eq:S1_offset} and the power-law pre-factors in Eq. \eqref{eq:Sq_offset} to vanish. 
Hence, the numerical values of $S_1^{(t)}$ and $S_q^{(t)}$ are free from discretization artifacts and are directly comparable across all chain lengths. 

To quantify the departures from the extensive Boltzmann--Gibbs limit, we define the per-frame ratio
\begin{equation}
	\left(\frac{S_q}{S_1}\right)^{\!(t)}
	= \frac{S_q^{(t)}}{S_1^{(t)}} \, 
	\label{eq:Sq_S1_ratio}
\end{equation}
which is analyzed as a function of the corresponding $q_{\text{fit}}^{(t)}$.
In the extensive limit ($q \to 1$), $S_q^{(t)} \to S_1^{(t)}$ and the ratio converges to unit. 
Deviations from unity quantify the influence of persistent conformational heterogeneities on non-Gaussian behavior, and their systematic evolution with chain length is analyzed in Sec. \ref{subsec:results:entropy-variation}.

\begin{figure*}
	\centering
	\includegraphics[width=0.95\textwidth]{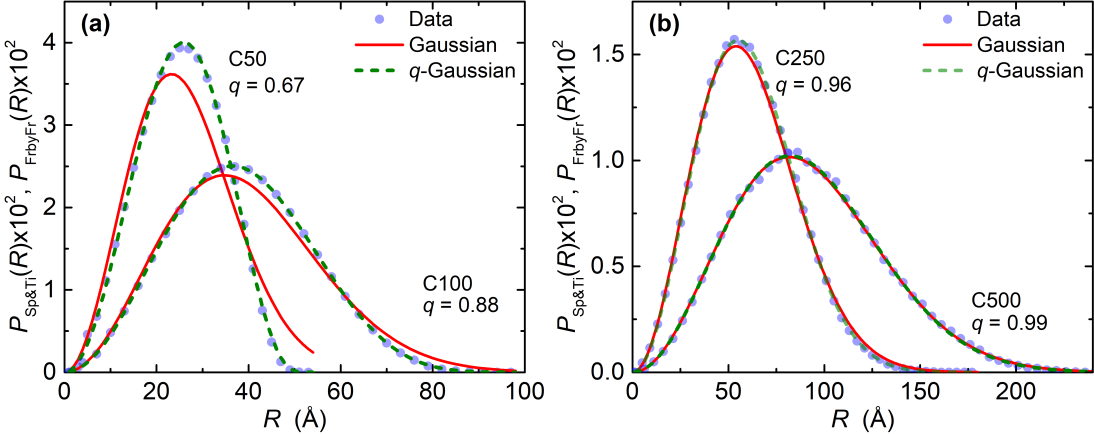}
	\caption{Data histogram for the end-to-end distances of the different polymer systems and fits with a Gaussian, Eq. (\ref{eq:probability-Gauss_radial}), and $q$-Gaussian, Eq. (\ref{eq:q-Gaussian-general form-radial}). (a) C50 and C100, (b) C250 and C500. The entropic index $q$ values are indicated. 
	The fit quality parameters are provided in Table~\ref{tab:fit-quality-main} 
	and supplementary material, Table~S1.
	} 
	\label{fig:fig-hist-gauss-q-gauss}
\end{figure*}

\begin{figure*}
	\centering
	\includegraphics[width=0.95\textwidth]{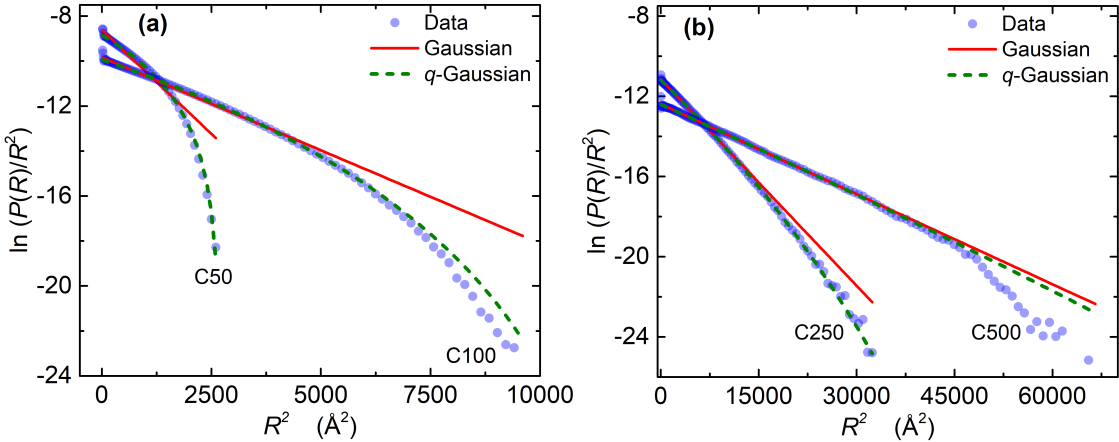}
	\caption{Linearity test in $R^2$. Plots of the logarithm of the reduced density as a function of $R^2$ and fits with the Gaussian (a) and \textit{q}-Gaussian (b) as described in Fig.  \ref{fig:fig-hist-gauss-q-gauss}. Eq. (\ref{eq:propoto-R^2}) describes the Gaussian fit (solid line) and Eq. (\ref{eq:propoto-R2-q-gaussian}) describes the $q$-Gaussian fit.  Notice the tail in all systems and superimposed fit with the $q$-Gaussian, except for the C500 system. Note that a plot of $\ln_q \left[ P_q(R)/R^2\right]$ \textit{vs.} $R^2$, should yield a straight line for C50--C250, according to Eq. (\ref{eq:propoto-R2-q-gaussian})--supplementary material, Fig. S2.
	} 
	\label{fig:fig-ln-P-R-R2}
\end{figure*}

\begin{figure*}
	\centering
	\includegraphics[width=\textwidth]{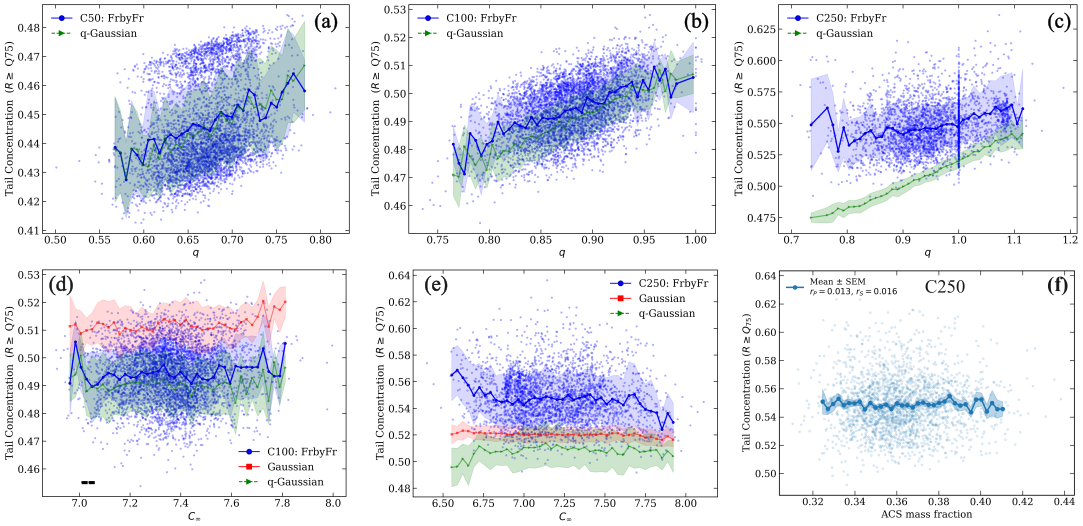}
	\caption{Tail concentration of the last data quartile ($R\ge$Q75) for the different polymer systems. 
	Blue symbols represent individual simulation snapshots (frame-by-frame, FrbyFr). 
	The solid blue line and shaded band show the binned mean $±\pm 1$ SEM. Panels (a), (b) and (c) represent the correlations between the tail concentration and $q$ for C50, C100 and C250, respectively. 
	Panels (d) and (e) illustrate the absence of correlation between the tail concentration and the characteristic ratio $C_\infty$ for C100 and C250, respectively. Panel (f) shows the analogous result for the ACS mass fraction, illustrated for C250;
	the Pearson and Spearman correlation coefficients 
	are $r_P = 0.013$ and $r_S = 0.016$, respectively, confirming the 
	absence of a linear or monotonic correlation (weak correlations requires $|r|>0.3$).
		} 
	\label{fig:fig-tails-time-q}
\end{figure*}

\section{Results and Discussion}
\subsection{\label{subsec:results:ETE-distribution}End-to-end distance distributions}

Figures \ref{fig:fig-hist-gauss-q-gauss}(a,b) show the normalized probability 
distributions, $P(R)$, for the end-to-end distances of all four polymer systems considered, from C50 up to C500. 
Data histograms obtained via space-time averaging (Sp\&Ti) and frame-by-frame averaging (FrbyFr) superimpose perfectly.
Consequently, fits for both averaging methods yield identical parameters within round-off errors.
Detailed fit quality results are summarized in Table \ref{tab:fit-quality-main} and Table S1 (supplementary material, Sec. S4). 

As expected, and as shown in Fig.  \ref{fig:fig-hist-gauss-q-gauss}(a), deviations from Gaussianity are more pronounced in the unentangled systems (C50 and C100). 
In these cases, the $q$-Gaussian accurately follows the data, with the entropic index increasing from $q = 0.67$ (C50) to $q = 0.88$ (C100). 
For the entangled systems (C250 and C500), the differences between the two models
fade as $q$ approaches unity [Fig. \ref{fig:fig-hist-gauss-q-gauss}(b)].
While the peak region for C250 and C500 appears nearly Gaussian, the $q$-Gaussian
remains necessary to describe the tail behavior, particularly for C250.

This behavior is more clearly visible in the $\ln[P(R)/R^2]$ \textit{vs.} $R^2$ representation (Fig.  \ref{fig:fig-ln-P-R-R2}).
Here, a true Gaussian appears as a straight line (Fig.~S2). 
However, the data shows a clear downward curvature at large $R^2$ for all chain lengths, with the deviation from linearity increasing as chain length decreases. 
The $q$-Gaussian [Eq.~(\ref{eq:propoto-R2-q-gaussian})] describes this curvature for C50, C100, and C250.
For C500, the $q$-Gaussian is nearly indistinguishable from the Gaussian, consistent with $q = 0.99$; both models slightly overestimate the tail at the highest $R^2$ values [Fig.~\ref{fig:fig-ln-P-R-R2}(b)].

The non-Gaussianity parameter $\alpha_{2}$ remains negative for all systems (Table~\ref{tab:fit-quality-main}), with its magnitude decreasing from $-0.153$ (C50) to $-0.026$ (C500).
The persistent negative sign is a physical requirement, reflecting the finite maximum chain extension (compact support) correctly captured by the $q$-Gaussian but neglected by the Gaussian. 
Because $\alpha_{2}$ is determined entirely from the data moments, independent of any fitting procedure, it provides definitive evidence that the underlying statistics are non-Gaussian.

Quantitative fit criteria (Table \ref{tab:fit quality criteria}) confirm the superiority of the $q$-Gaussian for C50, C100, and C250, demonstrating
further a systematic evolution of the entropic index from $q = 0.67 $(C50) to $q = 0.99 $(C500).
For C50, C100, and C250, the $q$-Gaussian is unambiguously favored: the residuals, 
$\chi^{2}/\nu$, $\alpha_2$, Q--Q plots, the skewness and the kurtosis (supplementary material, Sec. S4, Table S1) all point in the same direction. 
For C500, both models adequately describe the peak.
While the elevated $\chi^{2}/\nu$ in C500 originates from residual noise in the tail, the Q–Q plots and the sign of $\alpha_2$ continue to favor the $q$-Gaussian. 
However, since $q = 0.99$ is indistinguishable from unity within uncertainty, the $q$-Gaussian offers no significant practical advantage over the Gaussian for the longest chains. 
This result signals the effective masking of local Kuhn segment heterogeneities.

\begin{table}[ht]
	\centering
	\caption{%
	Fit quality criteria comparing Gaussian (\textbf{G}) and $q$-Gaussian (\textbf{$q$-G}) models.
	The entropic index $q$ evolves systematically from $q = 0.67$ (C50,	strongly non-Gaussian) to $q = 0.99$ (C500, nearly Gaussian).
	The non-Gaussianity parameter $\alpha_2 = \tfrac{3}{5}\langle R^4\rangle/\langle R^2\rangle^2 - 1$	remains persistently negative, indicating compact support (deficit of extended configurations).
	Fitting parameters in supplementary material, Table S2.
	}
	\label{tab:fit-quality-main}
\begin{tabular}{lcccc}
	\hline\hline
	Metric & C50 & C100 & C250 & C500 \\
	\hline
	$q$                              & $0.671$ & $0.880$ & $0.956$ & $0.987$ \\
	$\alpha_2$ (data)                & $-0.153$ & $-0.087$ & $-0.037$ & $-0.026$ \\
	\hline
	$\chi^2/\nu$ (\textbf{G})        & $8.0\times10^{6}$ & $2.3\times10^{4}$ & $63$ & $42$ \\
	$\chi^2/\nu$ (\textbf{$q$-G})   & $589$  & $58$ & $3.0$ & $22$ \\
	\hline
	RMSE ($10^{-4}$\,\AA$^{-1}$) (\textbf{G})
	& $36.6$ & $9.28$ & $2.22$ & $1.14$ \\
	RMSE ($10^{-4}$\,\AA$^{-1}$) (\textbf{$q$-G})
	& $4.46$ & $0.97$ & $0.77$ & $1.06$ \\
	\hline
	$\mathrm{RMSE}_{\mathrm{QQ}}$ (\AA) (\textbf{G})
	& $1.72$ & $1.69$ & $1.14$ & $0.81$ \\
	$\mathrm{RMSE}_{\mathrm{QQ}}$ (\AA) (\textbf{$q$-G})
	& $0.16$ & $0.12$ & $0.07$ & $0.39$ \\
	\hline\hline
\end{tabular}
\end{table}

\begin{figure*}
	\centering
	\includegraphics[width=0.80\textwidth]{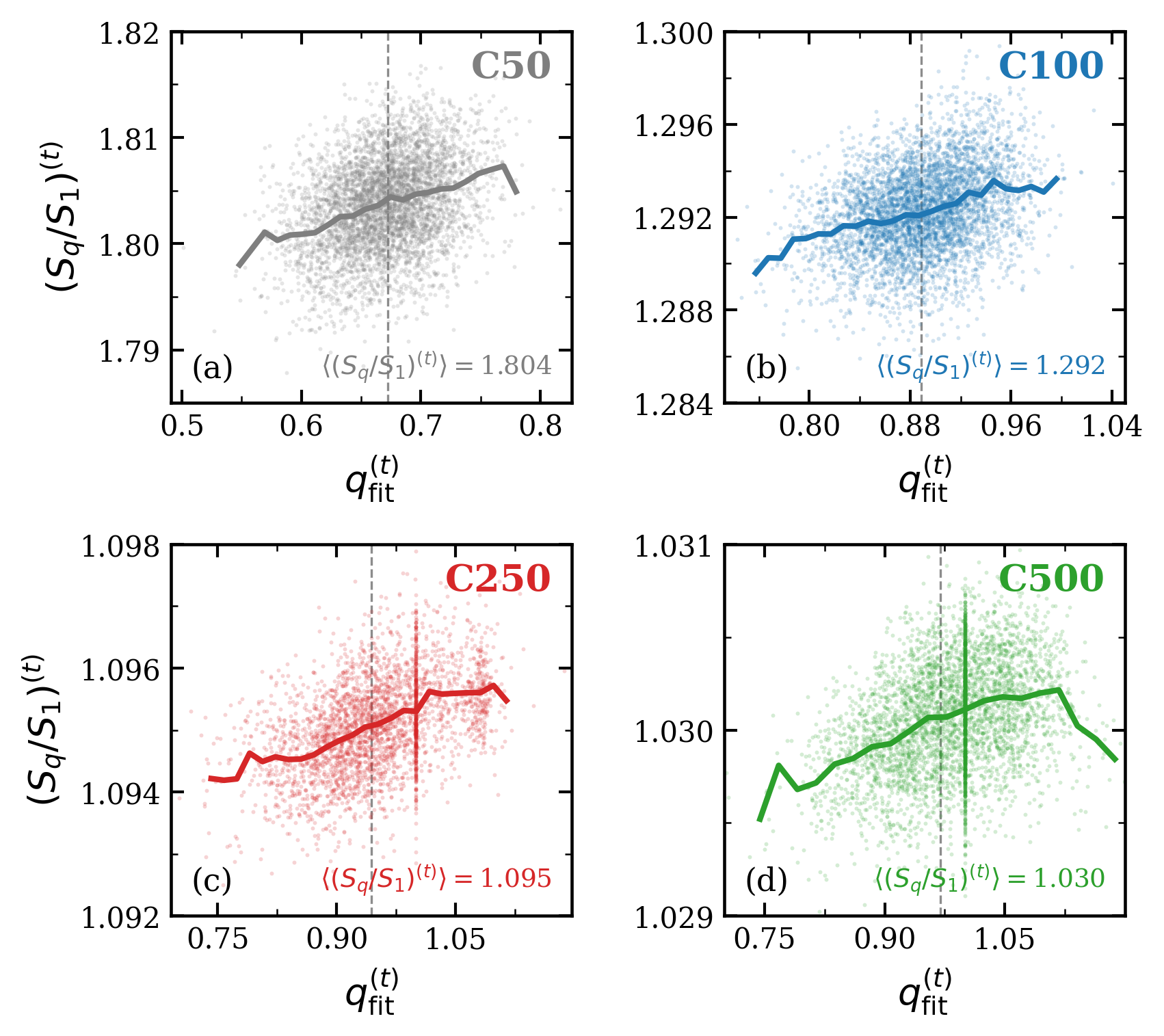}
\caption{
	Entropy non-extensivity quantified by the ratio $\bigl(S_q/S_1\bigr)^{(t)}$
	as a function of the entropic index $q_{\mathrm{fit}}^{(t)}$  for:
	(a)~C50; (b)~C100; (c)~C250; and (d)~C500.
	The time-weighted mean $\langle(S_q/S_1)^{(t)}\rangle$ is indicated
	in each panel.
	As in Fig.~\ref{fig:fig-tails-time-q}, each data point represents an individual simulation snapshot. Their running mean is represented by the solid line.
	The dashed vertical line marks the time-weighted mean $\langle q_{\mathrm{fit}}^{(t)}\rangle$ for each system, which agrees with the Sp\&Ti fitted value (Table~\ref{tab:fit-quality-main}) to within $2\%$.
	}
	\label{fig:fig-Sq-S1-vs-q}
\end{figure*}

\subsection{\label{subsec:results:tail-concentration}Tail Concentration}

Time-resolved plots of the last data quartile ($R \ge Q_{75}$) in the supplementary material (Sec. S5.3, Fig.~S3) demonstrate that the trajectories are stationary, confirming that the systems are well-equilibrated. 
These plots further illustrate how the $q$-Gaussian and Gaussian models follow the fluctuations observed in the data.

The superiority of the $q$-Gaussian fit in the tail is evident for C50 and C100 [Figs. S3(a,b)]. 
The validity limit of the $q$-Gaussian for describing extreme tail behavior is set by the C250 data in Fig. S3(c).

As described in Sec.~\ref{subsec:methods:tail}, the tail concentration $T_2$, defined by Eq.~\eqref{eq:T2-definition}, quantifies the fraction of the mean-square end-to-end distance $\langle R^2\rangle$ arising from configurations in the last quartile ($R \ge R_{Q_{75}}$).
For a theoretical Gaussian distribution, $T_2 = 53.4\%$.%
\footnote{For the normalized 1D radial Gaussian probability density 
	$P(x) = \sqrt{54/\pi}\,x^2\,e^{-3x^2/2}$, where $x = R/\sqrt{\langle R^2\rangle}$, 
	the 75th percentile is $x_{Q_{75}} \approx 1.1702$.
	The integrals in the tail concentration [Eq.~(\ref{eq:T2-definition})] are of the type $\int_{u}^{\infty} x^4 e^{-a x^2} dx$. The integral in the numerator (with $u = \mathrm{Q_{75}}$) is the upper incomplete Gamma function; that in the denominator (with $u = 0$) is the complete Gamma function. Their ratio is 
	$T_2 = \Gamma\!\left(\tfrac{5}{2},\,\tfrac{3}{2}x_{Q_{75}}^2\right)/\Gamma\!\left(\tfrac{5}{2}\right) = 0.534$, 
	where $\Gamma(s,z)$ is the upper incomplete gamma function.}
Thus, even in the absence of non-Gaussianity, the most extended 25\% of chains account for more than half of the total $\langle R^2\rangle$.

Observed values vary from $T_2 \approx 45\%$ for C50 to $\approx 55\%$ for C250 and C500. The value for C50 falls significantly below the Gaussian baseline, a direct consequence of the compact support inherent to the $q$-Gaussian at $q = 0.67$. 
This bounded support suppresses the contribution of the most extended configurations to $\langle R^2\rangle$ relative to the Gaussian prediction.
As $q$ increases with chain length, $T_2$ recovers toward the Gaussian 
baseline, approaching it for the entangled C250 and C500 systems.

To explain these results, we examined correlations between $T_2$, the entropic index $q$, the characteristic ratio $C_\infty$, and the ACS mass fraction. Fluctuations in $q$ [Figs.~\ref{fig:fig-tails-time-q}(a--c)] behave as expected: the slopes of $T_2$ versus $q$ decrease as Gaussianity is recovered, approaching a horizontal line as $q \to 1$.
Results for $C_\infty$ and the ACS mass fraction are presented in Figs.~\ref{fig:fig-tails-time-q}(d--f).
Intuitively, one might expect that the most extended configurations would result from an excess of ACS and these aligned chain segments would be associated with chains of relatively higher characteristic ratio.

However, these results demonstrate that $T_2$ is insensitive to 
fluctuations in both $C_\infty$ and the ACS mass fraction.
Thus, the origin of the most extended chain configurations that populate 
the tail of the end-to-end distance distribution must be reassessed.
Due to the absence of correlations with both quantities we conclude that these extended configurations should have a statistical origin.
Nevertheless, since $T_2$ approaches values consistent with Gaussianity 
for long chains, we may consider this a third independent criterion to 
infer the recovery of Gaussian statistics.

\subsection{\label{subsec:results:entropy-variation}Entropy Variation}

We analyze the entropy variation following a procedure similar to that employed for the tail concentration. 
To quantify the degree of extensivity, we use the per-frame ratio $\bigl(S_q/S_1\bigr)^{(t)}$. 
The time-resolved evolution of this quantity (supplementary material, Fig. S4) displays stationary fluctuations, further validating the FrbyFr construction and confirming that all systems are well equilibrated.

Besides equilibration, the most relevant feature of Fig. S4 is the quantification of non-extensivity.
As shown also in Fig.~\ref{fig:fig-Sq-S1-vs-q}, the non-extensivity is strongly chain-length dependent, decreasing systematically from C50 to C500. 
This behavior is consistent with the global and tail statistics presented in Figs.~\ref{fig:fig-hist-gauss-q-gauss} and~\ref{fig:fig-tails-time-q}.

The magnitude of these fluctuations is small compared to the systematic 
variation, indicating that the observed non-extensivity reflects an 
intrinsic property of the chain statistics. 
The superadditivity of the Tsallis entropy for short chains ($S_q > S_1$) 
indicates that these systems explore a broader range of conformational 
states than an extensive, Gaussian system would predict.

The FrbyFr time-weighted mean $\langle q_{\mathrm{fit}} \rangle$ agrees with the Sp\&Ti fitted values (Table~\ref{tab:fit-quality-main}) within 2\% for all systems, confirming the consistency of both averaging procedures.
Similarly to the tail-concentration results, the $S_q/S_1$ ratio is unaffected by fluctuations in either the characteristic ratio or the ACS mass fraction (supplementary material, Fig.~S5). 

Also consistent with the tail-concentration results, the slopes of $\bigl(S_q/S_1\bigr)^{(t)}$ versus $q_{\mathrm{fit}}$ decrease progressively from C50 to C500 [Figs.~\ref{fig:fig-Sq-S1-vs-q}(a--d)]. 
The entropy is clearly superadditive for C50 and C100, with $\langle(S_q/S_1)^{(t)}\rangle \approx 1.804$ and $1.292$, respectively [Figs.~\ref{fig:fig-Sq-S1-vs-q}(a,b)]. 
For the entangled C250 and C500 systems, the entropy remains weakly superadditive, with $\langle(S_q/S_1)^{(t)}\rangle \approx 1.095$ and $1.030$. 
In particular, for C500 [Fig.~\ref{fig:fig-Sq-S1-vs-q}(d)], the per-frame entropic index $q_{\mathrm{fit}}$ fluctuates across unity. 
This indicates that, as a result of thermal fluctuations in the end-to-end distance distribution, individual configurations sample both the superadditive ($q < 1$) and subadditive ($q > 1$) regimes, the latter corresponding to heavier tails than the Gaussian, signaling the recovery of extensive Boltzmann--Gibbs statistics.

\begin{figure*}
	\centering
	\includegraphics[width=0.95\textwidth]{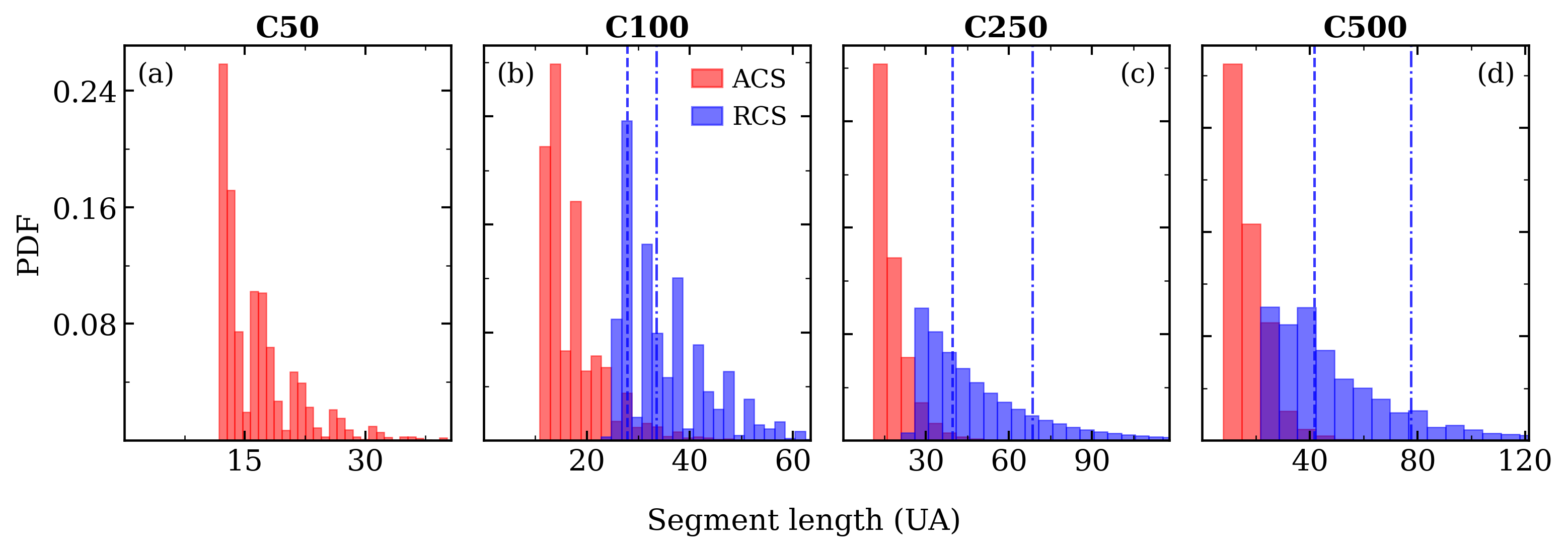}
	\caption{
	Segment length distributions in terms of united atoms (UA) for ACS (red) and RCS (blue) in all four systems.
	C50 has no RCS segments. 
	Only RCS segments containing at least two Kuhn segments ($n_{\rm seg} \geq 2\,n_k$) are included. 
	The $x$-axis in UA units is directly proportional to the molar mass of the segment ($M = n_{\rm UA} \times 14.03$~\si{g.mol^{-1}}~for polyethylene), so the distributions are the same as those in Fig. {S6} in the supplementary material, expressed in different units. 
	The vertical dashed and dash-dotted lines mark $M_{n,\mathrm{RCS}}$ and $M_{z,\mathrm{RCS}}$ from Table~\ref{tab:substructure_data}, converted to UA units. The distributions decay following a Flory-like form, with an increasing number of RCS segments per chain as chain length increases.
	}
	\label{fig:fig_segment_length_hist}
\end{figure*}
%
\begin{figure*}
	\centering
	\includegraphics[width=0.95\textwidth]{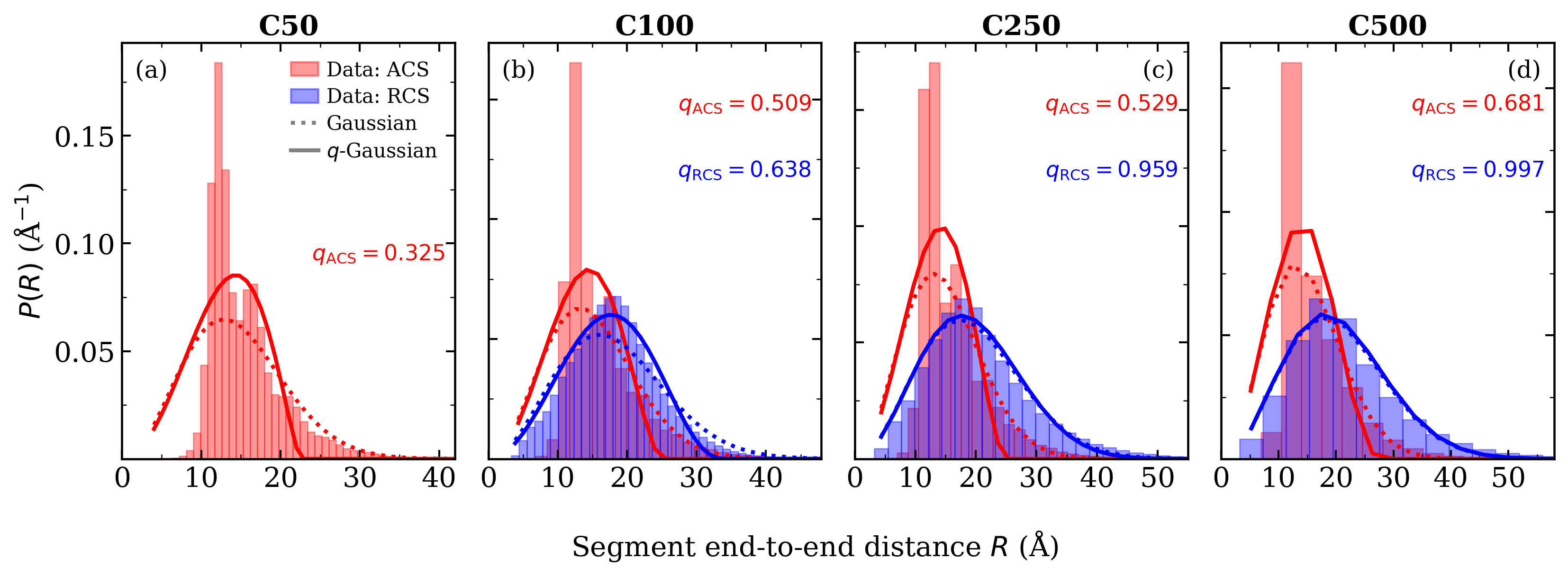}
	\caption{
	End-to-end distance distributions $P(R)$ for ACS (red bars)	and RCS (blue bars) segments for all four systems.
	(a)~C50: ACS only ($q_{\rm ACS} = 0.325$); no RCS segments are present.
	(b)~C100: ACS ($q_{\rm ACS} = 0.509$) and RCS ($q_{\rm RCS} = 0.638$); both are non-Gaussian, with RCS exhibiting intermediate behavior.
	(c)~C250: ACS ($q_{\rm ACS} = 0.529$) remain strongly non-Gaussian, while RCS ($q_{\rm RCS} = 0.956$) are nearly Gaussian, consistent with the whole-chain fit value $q = 0.956$ (Table~\ref{tab:fit-quality-main}).
	(d)~C500: ACS ($q_{\rm ACS} = 0.681$) continue to exhibit non-Gaussian 	statistics, while RCS ($q_{\rm RCS} = 0.997$) are indistinguishable from Gaussian, consistent with $q = 0.987$ for the whole chain.
	In each panel, the dotted lines are Gaussian fits and the solid lines are $q$-Gaussian fits (red for ACS, blue for RCS).
	Fit quality criteria are reported in Table~S2 of the supplementary material.
	}
	\label{fig:fig_ACS_RCS_fits}
\end{figure*}

\begin{figure*}
	\centering
	\includegraphics[width=0.95\textwidth]{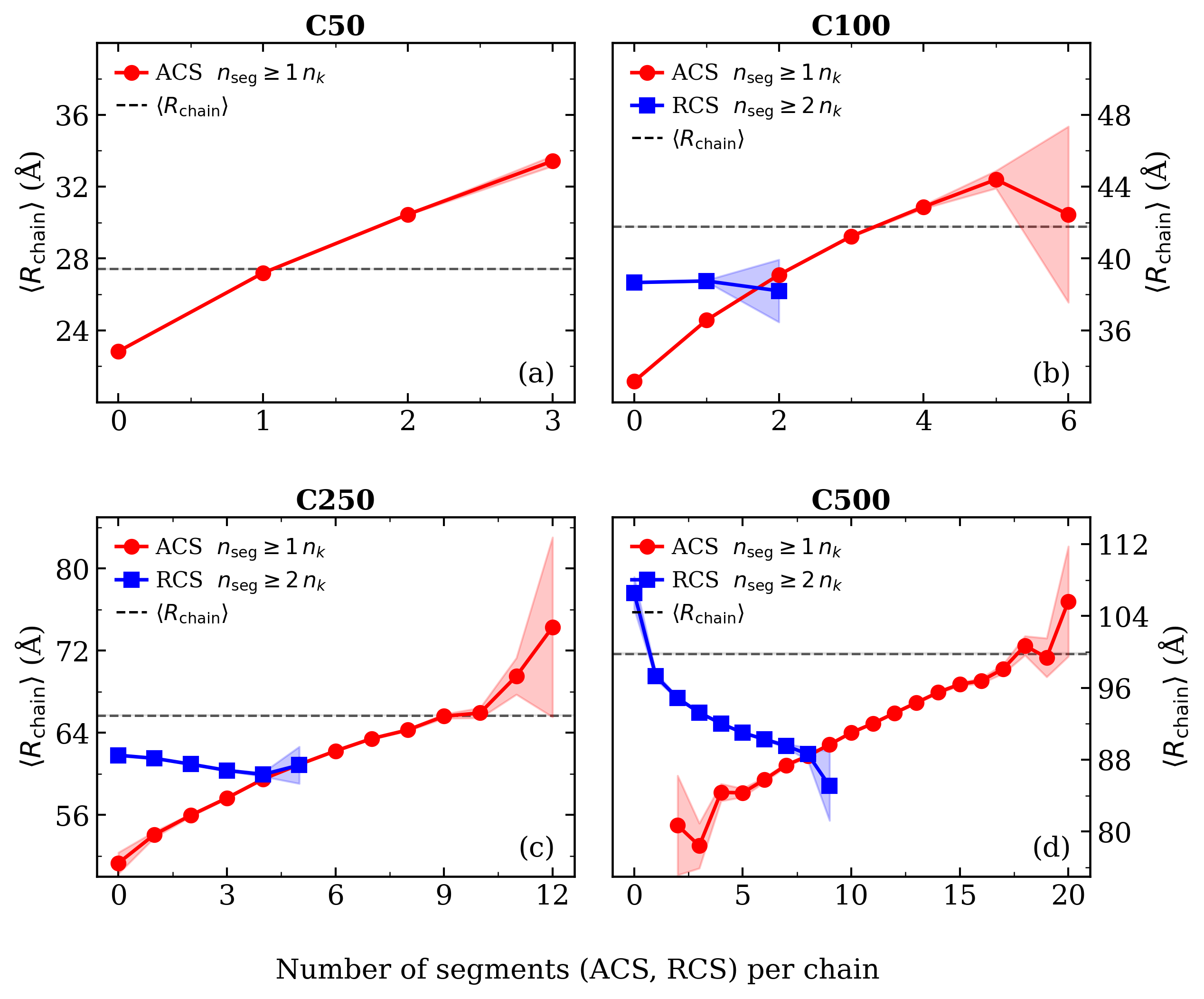}
	\caption{
		Correlation between the average end-to-end distance of the
		chain and the number of ACS and RCS segments present in each
		chain.
		C50, in panel (a) has no segments in random conformational
		sequences; only ACS are present.
		C100 has up to 6 Kuhn segments in ACS per chain, being these
		the most extended chains, and has two segments in RCS only,
		each with $\approx2.65$~Kuhn segments
		(Table~\ref{tab:substructure_data}).
		C250 has a maximum of 12 ACS and 5 segments in RCS, with a
		number average molar mass of $\approx555$~\si{g.mol^{-1}}.
		C500 has a maximum of 9 segments in RCS and a maximum of
		20 ACS per chain.
	}
	\label{fig:fig-R2-n-ACS-n-RCS-correlations}
\end{figure*}

\subsection{\label{subsec:results:physical-origin}Physical origin 
	of non-Gaussianity and crossover to Gaussianity}

As discussed in Sec.~\ref{sec:introduction}, polymer chains exhibit intrinsic conformational heterogeneity at the scale of the Kuhn segment. We identified previously three distinct types of Kuhn-scale substructures: aligned chain segments (ACS), random conformational sequences (RCS), and chain ends (CE).\cite{Martins-macromol-2013,Martins-2026}
Because a single Kuhn segment is statistical but non-Gaussian, and the smallest entropic spring with mild non-Gaussianity contains two Kuhn segments,\cite{Martins-2026} the minimal size of RCS was set to this threshold.
The segment length distributions for ACS and RCS in all four systems are shown in Fig.~\ref{fig:fig_segment_length_hist}.
The longest ACS contain $\approx50$\,UA (4--5 Kuhn segments), while the longest RCS contain $\approx125$\,UA (10--11 Kuhn segments). C50 has no RCS segments [Fig.~\ref{fig:fig_segment_length_hist}(a)].

ACS are characterized by local alignment and reduced conformational freedom; they interact with one another to form short-range ordered regions in polymer melts,\cite{Martins-macromol-2013} with sizes comparable to those reported for dynamic heterogeneities in supercooled melts ($\approx2$~nm).\cite{Ediger-2003,Ediger-annualreviews-2000}
These interactions are both intramolecular, arising from bond-angle and torsional energy barriers, and intermolecular, resulting primarily from van der Waals interactions between
aligned segments. The interaction energy between two parallel Kuhn segments in polyethylene is $\approx1$~\si{kJ.mol^{-1}} and increases with the number of interacting segments due to their additive nature.\cite{Martins-2011} 
Consequently, ACS exhibit orientational relaxation times roughly an order of magnitude longer than those of RCS and CE, with stretched-exponential relaxation characterized by $\beta\approx 0.5$, consistent with locally constrained, quasi-one-dimensional
defect diffusion. 
In contrast, the RCS segments contain a comparatively larger number of united atoms, and a smaller persistent length resulting from their folded character
($l_{\text{p,ACS}}\approx15$\,\AA\ and $l_{\text{p,RCS}}\approx10$\,\AA; 
Fig.~7(d) in Ref.~\onlinecite{Martins-2026})
which is reflected in negative values for the bond-vector correlation function [Fig. 7(a) in Ref.~\onlinecite{Martins-2026}]. 
The coiling of the RCS segments is evident in their end-to-end distance values displayed in Fig.~\ref{fig:fig_ACS_RCS_fits}(b-d). 

As shown in Figs.~\ref{fig:fig_ACS_RCS_fits}(a--d), the ACS end-to-end distance distributions are markedly non-Gaussian across all chain lengths. 
Neither the Gaussian nor the $q$-Gaussian models provide an adequate description for $\langle R_\text{ACS} \rangle$. Specifically, the $q$-Gaussian  yields $q$ values   with large relative uncertainties, ranging from 92\% to 31\% (supplementary material, Sec. S4.2, Table S2).
This failure is an intrinsic consequence of ACS structural rigidity due to the extended orientational correlations, intermolecular interactions between neighboring ACS segments, and slow relaxation.  
Also, because most ACS contain only a single Kuhn segment, their statistical degrees of freedom are severely limited. 
Consequently, the $q_{\mathrm{ACS}}$ values are less reliable than the corresponding $q_{\mathrm{RCS}}$ parameter for the $RCS$ end-to-end distance distribution, $\langle R_\text{RCS} \rangle$. 
The RCS distributions evolve systematically: for C100 they are moderately
non-Gaussian ($q_{\mathrm{RCS}}=0.638$), for C250 they are nearly Gaussian
($q_{\mathrm{RCS}}=0.956$), and for C500 they are indistinguishable from Gaussian
($q_{\mathrm{RCS}}=0.997$). Here the uncertainties in the $q$  values vary from $\approx 7\%$ to $\approx 3\%$.

Figure \ref{fig:fig-R2-n-ACS-n-RCS-correlations} shows that ACS and RCS have
divergent  effects on the chain end-to-end distance: more ACS per chain increases
$\langle R_{\mathrm{chain}}\rangle$ in all systems, while more RCS reduces it,
most clearly in C500 [Fig. \ref{fig:fig-R2-n-ACS-n-RCS-correlations}(d)].
Where ACS and RCS are present in equal numbers --- two of each in C100, four to
five in C250, and eight in C500 --- the resulting $\langle R_{\mathrm{chain}}\rangle$
is consistently below the overall average $\langle R \rangle$, indicating that
the folding effect of RCS per segment exceeds the extension effect of ACS.

In C50 [Fig.~\ref{fig:fig-R2-n-ACS-n-RCS-correlations}(a)], RCS are entirely absent, and 
the chain end-to-end distance is governed solely by ACS and chain ends.
Non-Gaussianity and non-extensivity appear to be correlated: the non-Gaussian character of the ACS distribution in C50 yields $q = 0.67$, $T_2\approx 45\%$ and $S_q/S_1 = 1.80$.
The observed non-Gaussianity does not correlate with the instantaneous ACS mass fraction, the characteristic ratio, or the number of aligned segments alone [Figs.~S3(d--f) and Figs.~S5(a--d), supplementary material, Sec. S5].
Instead, it originates from the heterogeneous chain architecture at the Kuhn-segment level.

As RCS accumulate with increasing chain length (C100, C250 and C500), their nearly Gaussian distributions [Fig.~\ref{fig:fig-R2-n-ACS-n-RCS-correlations}(b--d)] progressively dominate the whole-chain $P(R)$, driving $q \to 1$. Furthermore, the $q_\text{RCS}$ values compare to those obtained for the whole chain [Fig.~\ref{fig:fig-hist-gauss-q-gauss}(a-b)].  
Most important, the number-average molar mass of RCS increases with chain length and 
saturates at $M_{n,\mathrm{RCS}}/M_k \approx 3.6$ (Table~\ref{tab:substructure_data}), at the same chain lengths where the whole-chain distribution is well described by a Gaussian.

The crossover to Gaussianity is therefore neither abrupt nor controlled by a single structural parameter.
It reflects the emergence of a sufficient number of Gaussian RCS domains along the chain contour to compensate  the intrinsically non-Gaussian statistics of ACS, a mechanism quantified by $q \to 1$ and $S_q/S_1 \to 1$ simultaneously.
The ratio $S_q/S_1$, computed directly from simulation data without any fitting,
decreases accordingly from $1.80$ (C50) to $1.03$ (C500), tracking this structural evolution quantitatively.
Even when this averaging occurs, ACS persist in the molten state: their presence and effect on relaxation dynamics cannot be neglected even when obscured at the full-chain scale.

We conclude then that the Gaussian recovery is a statistical masking effect: it is not
reflected in $\langle R^2 \rangle$, which scales normally throughout and is insensitive to the redistribution of statistical weight between ACS-dominated and RCS-dominated chain configurations.

\begin{figure}[h!]
	\centering
	\includegraphics[width=0.48\textwidth]{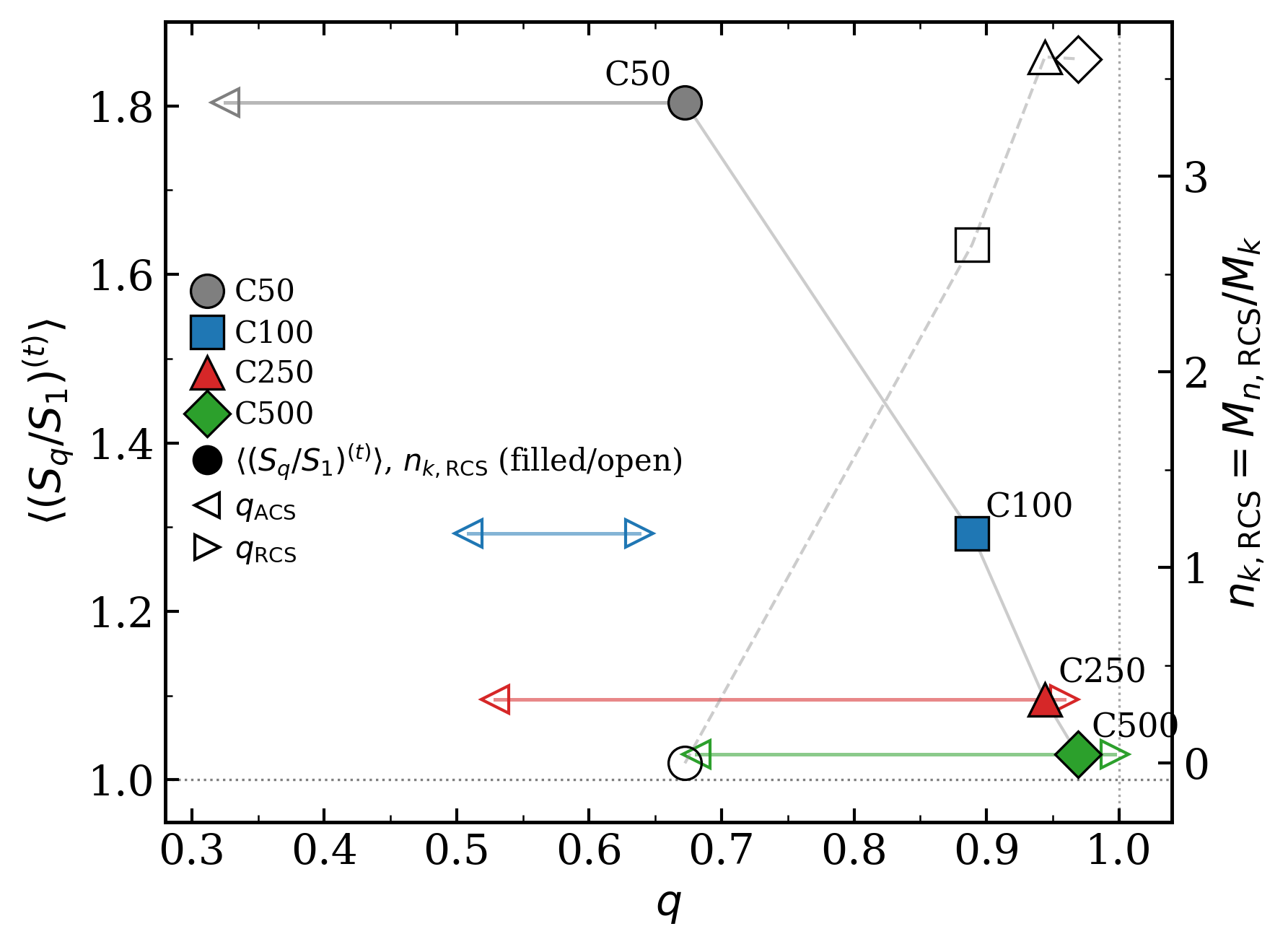}
	\caption{
		Time-weighted mean entropy ratio
		$\langle(S_q/S_1)^{(t)}\rangle$ (filled symbols, left axis)
		and number of Kuhn segments per RCS,
		$n_{k,\mathrm{RCS}}=M_{n,\mathrm{RCS}}/M_k$ (open symbols,
		right axis), plotted against the entropic index $q$ for each
		system.
		Three $q$ values are shown at the
		$\langle(S_q/S_1)^{(t)}\rangle$ level of each chain:
		(i)~the time-weighted mean whole-chain value
		$\langle q_{\mathrm{fit}}^{(t)}\rangle$ (filled marker),
		(ii)~the $q$-Gaussian fit to the ACS distribution
		$q_{\mathrm{ACS}}$ (open left-pointing triangle), and
		(iii)~the $q$-Gaussian fit to the RCS distribution
		$q_{\mathrm{RCS}}$ (open right-pointing triangle; absent
		for C50, which has no RCS segments).
		The horizontal line connecting the three points for each
		chain spans the heterogeneity in $q$ at that level of
		entropy non-extensivity.
		While $q_{\mathrm{RCS}}$ converges rapidly toward unity
		from C100 to C500, $q_{\mathrm{ACS}}$ remains substantially
		below unity for all systems, and the whole-chain $q$
		converges to $q_{\mathrm{RCS}}$, reflecting the dominant
		contribution of RCS to the overall chain statistics and
		illustrating the statistical masking mechanism.
	}
	\label{fig:Sq-S1-and-nK-RCS-vs-q}
\end{figure}

Figure~\ref{fig:Sq-S1-and-nK-RCS-vs-q} synthesizes this evolution by  displaying, for each system, the time-weighted mean entropy ratio $\langle(S_q/S_1)^{(t)}\rangle$, the number of Kuhn segments per RCS  ($n_{k,\mathrm{RCS}}$), and the whole-chain entropic index  $\langle q_{\mathrm{fit}}^{(t)}\rangle$ together with $q_{\mathrm{RCS}}$, 
against the entropic index $q$.
For C50, RCS are entirely absent and the whole-chain $q = 0.67$ reflects  exclusively the non-Gaussian character of ACS and chain ends.
For C100, RCS begin to appear but remain scarce and moderately non-Gaussian ($q_{\mathrm{RCS}} = 0.638$), while the whole-chain $q = 0.880$ already exceeds $q_{\mathrm{RCS}}$, indicating that chain ends contribute to the partial recovery of Gaussianity.
The qualitative change occurs at C250: $q_{\mathrm{RCS}} = 0.959$ is nearly indistinguishable from the whole-chain fitted value $q = 0.956$, signaling that RCS now dominate the global chain statistics.
For C500, $q_{\mathrm{RCS}} = 0.997$ is indistinguishable from Gaussian, consistent with the whole-chain $q = 0.987$.
The convergence of the whole-chain $q$ toward $q_{\mathrm{RCS}}$ directly illustrates the statistical masking mechanism: as RCS accumulate, they dominate the global chain statistics and progressively obscure the non-Gaussian character of the ACS domains 	
(Sec.~\ref{subsec:results:physical-origin}, Fig.~\ref{fig:fig_ACS_RCS_fits}).

\begin{table}[h!]
	\centering
	\caption{
		Characterization of the conformational substructure in
		polyethylene melts. $M_k$ is the Kuhn segment molar mass;
		$M_{n,RCS}$ and $M_{z,RCS}$ are the number and $z$-average
		molar masses for RCS segments, respectively, evaluated from
		data in Fig.~S6 of the supplementary material (or
		equivalently from Fig.~\ref{fig:fig_segment_length_hist}).
		The ratio $M_{n,RCS}/M_k$ quantifies the entropic spring.
		ACS-M$_f$ and LOR-M$_f$ are the mass fractions of aligned
		chain segments and local ordered regions, respectively.
		All molar mass values, expressed in \si{g.mol^{-1}},
		represent mean values with a standard error of less than
		10\%.
	}
	\label{tab:substructure_data}
	\begin{ruledtabular}
		\begin{tabular}{l c S[table-format=3.2] S[table-format=1.2]
				S[table-format=4.2] S[table-format=1.3] S[table-format=1.3]}
			System & {$M_k$} & {$M_{n,RCS}$} & {$M_{n,RCS}/M_k$}
			& {$M_{z,RCS}$} & {ACS-M$_f$} & {LOR-M$_f$} \\
			\midrule
			C50  & 136.12 & {---}  & {---} & {---}   & 0.234 & 0.077 \\
			C100 & 147.25 & 390.40 & 2.65  & 470.53  & 0.327 & 0.172 \\
			C250 & 153.85 & 555.44 & 3.61  & 959.54  & 0.361 & 0.137 \\
			C500 & 162.34 & 584.91 & 3.60  & 1087.10 & 0.382 & 0.096 \\
		\end{tabular}
	\end{ruledtabular}
\end{table}

\section{Conclusions}

We have provided a deeper insight into the journey of a polymer chain toward Gaussian statistics. 
The determining factor is the number and length of segments in random conformational sequences (RCS) connecting aligned chain segments (ACS). 
This evolution is governed by a competition between local structural constraints and global self-averaging capacity.

The end-to-end distance distributions of polyethylene melts, from the unentangled (C50, C100) to the entangled (C250, C500) regimes, are accurately described by a $q$-Gaussian distribution.
The entropic index $q$ increases systematically from $0.67$ to $0.99$, providing a unified framework to link Kuhn-scale heterogeneity, non-Gaussianity, and non-extensivity. 
We identify $q$ as a \textit{heterogeneity index}, tracking the structural evolution of the chain from a non-Gaussian, non-extensive state when RCS are absent to the Gaussian, extensive limit as the chain contour accumulates several RCS segments.

Our results establish that the physical origin of non-extensive behavior in short chains is a structural deficit: the absence of sufficient random conformational sequences to satisfy the requirements for statistical independence. 
The Central Limit Theorem (CLT) requires the sum of many independent variables to reach a Gaussian distribution. 
In short chains such as C50, the structural ``variables'' are restricted to ACS and CE, with a significant time-scale separation in their orientational relaxation ($\tau_{ACS} \approx 10 \tau_{CE}$). 
In this regime, the chain lacks the statistically independent RCS units that act as ``entropy reservoirs'', providing the conformational diversity required for the onset of extensivity.
Because the $q$ values of the RCS segments approach those of the whole chain as Gaussianity and extensivity are recovered ($S_q/S_1 \to 1$), we identify these segments as the structural units that provide the statistical independence required for the onset of extensivity.

Consequently, in the non-extensive regime, the chain contour is a mosaic of slow-relaxing ACS with constrained 1D diffusion and highly mobile chain ends (CE).
This configuration results in a highly superadditive entropy ratio ($S_q/S_1 \approx 1.8$), indicating that the chain explores a correlated phase space whose statistical complexity exceeds that captured by Boltzmann--Gibbs (BG) statistics.
The microscopic origin of this statistical complexity -- specifically the apparent entropy paradox where shorter chains explore a richer configurational phase space than that allowed by BG statistics -- will be explored in future work. 
We tentatively attribute this superadditivity to the enhanced  conformational freedom of chain ends, whose contribution relative  to the total chain length is greatest in short chains.
We hypothesize that suppressing chain-end conformational freedom by increasing backbone rigidity would maintain non-Gaussianity but result in a phase space poorer than that allowed by BG statistics, manifesting as $q > 1$ and a subadditive entropy ratio ($S_q/S_1 < 1$).
This hypothesis could be tested in short semiflexible chains, or in backbone-branched systems where pendant groups of one to two Kuhn segments are inserted at regular intervals of two backbone Kuhn segments, structurally suppressing ACS formation.

Critically, the recovery of Gaussian statistics ($q \to 1$) must be interpreted as a \textit{statistical masking effect} rather than a structural dissolution of local order. 
Although ACS remain physically present across all chain lengths (constituting $\approx 35\%$ of the mass even at C500), their contribution to the global end-to-end vector is progressively obscured by the increasing population of statistically independent RCS units.
We identify the onset of this internal self-averaging at C100, where the chain first contains approximately two RCS segments ($M_{n,\mathrm{RCS}} \approx 2.65\,M_k$).

The RCS segments, besides being responsible for the emergence of Gaussianity, present other features worth exploring in detail. As shown in a previous work,\cite{Martins-2026} their translational dynamics are similar to those of segments at the middle of the chain, and the estimated value $n_\mathrm{RCS} k_\mathrm{B} T \approx 1.67$~MPa is comparable to the plateau modulus of polyethylene, suggesting a connection between RCS dynamics and entropic elasticity that deserves further investigation.

The persistence of local conformational constraints under an apparently  simple global behavior is not unique to polymer melts.
\citeauthor{Frauenfelder-1991} established that proteins populate a hierarchy of conformational substates with distinct structural and dynamic properties, which remain physically active even when macroscopic measurements such as ligand rebinding kinetics appear to follow simple behavior.\cite{Frauenfelder-1991}
The masking effect documented here is the polymer analogue: ACS domains persist and retain their structural and dynamic identity across all chain lengths studied, but their contribution to the global end-to-end distance distribution is progressively obscured by the accumulation of RCS segments.
In both cases, a global observable --- $R_g$ for polymers, ligand rebinding kinetics for proteins --- fails to capture the richer underlying heterogeneity that becomes visible only when the full distribution is examined.

Although the mean-square end-to-end distance scales linearly with chain length, 
$\langle R^2\rangle \propto N$, throughout all systems studied, including the non-Gaussian unentangled chains,\cite{Martins-macromol-2013} this single parameter is insensitive to the Kuhn-scale heterogeneity and non-Gaussianity. 
This has a direct consequence for the interpretation of SANS experiments: SANS measures $R_g$, which is sensitive only to $\langle R^2\rangle$, and is therefore blind to the transition from non-Gaussian to Gaussian statistics documented here.
The consistency of SANS with Gaussian scaling\cite{Benoit-1973,Benoit-1974,Schelten-1974} 
and with Flory's ideality hypothesis does not imply the absence of conformational heterogeneity at the Kuhn scale, nor the absence of non-Gaussian statistics in short and intermediate chains: it reflects the insensitivity of the second moment to the shape of $P(R)$.

Finally, we propose a \textit{unified statistical model} for non-Gaussian behavior in polymer melts based on a fundamental symmetry around the $q=1$ Gaussian fixed point.
While linear polyethylene demonstrates $q < 1$ due to the correlated mosaic of constrained ACS and mobile ends, a complementary regime with $q > 1$ is predicted for systems where ACS formation is suppressed and backbone stiffening dominates, such as semiflexible chains or specific branched architectures.
Confirming this symmetry and exploring its implications for chain  dynamics --- specifically the role of non-Gaussian equilibrium statistics in determining the entropic spring constant and relaxation times in unentangled and entangled systems --- are open issues worth exploring.

\section*{Supplementary Material}

The supplementary material provides the following: S1. Further details on the $q$-Gaussian, S1.1. On the concept of compact support in the q-Gaussian distribution, S1.2. The $q>1$ case: normalization and second moment; S2. Numerical procedures and definitions, S2.1. Data histogram and radial probability density normalization, S2.2. Models and fits, S2.3. Entropy discretization; S3. Discretization offsets in entropy definitions, S3.1. Boltzmann-Gibbs entropy ($S_1$), S3.2. Tsallis entropy ($S_q$), S3.3. The unit bin-width; S4. Fit quality criteria, S4.1. Fit-quality table: end-to-end distances of the chain, S4.2. Fit-quality table: end-to-end distances of ACS and RCS; S5. Complementary data results, S5.1. Q-Q plots, S5.2. Data linearization, S5.3. Tail concentration, S5.4. Entropy extensivity and correlations, S5.5. Molar mass distribution of the RCS segments. 

\section*{Acknowledgments}
I thank Loic Hillou and Sacha Mould for discussions.

\section*{AUTHOR DECLARATIONS}
\subsection*{Conflict of Interest}
The author has no conflicts to disclose.

\section*{Data Availability}
The data that support the findings of this study are available within the article and its supplementary material and are available from the author upon reasonable request.

\appendix
\section{Derivations for the 3D radial $q$-Gaussian}
\label{app:qgauss-derivations}

We provide here the intermediate mathematical steps supporting the expressions for the normalized 3D radial $q$-Gaussian probability density and its moments in the compact-support regime ($q<1$) presented in Sec.~ \ref{subsec:theory-q-gaussian}. For completeness we provide also the relevant equations for the $q>1$ case. Additional details on these last equations are  provided in the supplementary material, Sec.~ 1.2. 

\subsection{Normalization of the partition function $Z_q$}
The radial $q$-Gaussian probability density in three dimensions may be written as
\begin{equation} 
	\label{eqApp:radial-q-Gaussian-Zq}
	P_q(R)=\frac{4\pi R^2}{Z_q}\exp_q(-\beta_q R^2)\, ,
\end{equation}
where $Z_q$ ensures $\int_0^\infty P_q(R)\, dR=1$. For $q<1$ the distribution has compact support (supplementary material, {Sec.~ S1.1}) and the upper integration limit becomes $R_{\max}=1/\sqrt{\beta_q(1-q)}$, yielding
\begin{equation}
	Z_q^{(q<1)}=4\pi\int_{0}^{R_{\max}}R^2\left[\,1-(1-q)\beta_qR^2\,\right]^{\nu}\, dR,
	\label{eqApp:Zq-start}
\end{equation}
with $\nu=1/(1-q)$.

The integration of this equation follows the steps below
\begin{subequations}
	\begin{align}
		Z_q^{(q<1)} &=  \frac{2 \pi}{A ^{3/2}} \int_{0}^{1}  t^{1/2} \left[1 - t \right]^\nu dt \, , \label{eqApp:two} \\
		Z_q^{(q<1)} &=  \frac{ 2\pi}{A ^{3/2}} B\left(\frac{3}{2}, \nu-1\right) \, , \label{eqApp:three}\\
		Z_q^{(q<1)} &=  \frac{ 2\pi}{A ^{3/2}} \frac{\Gamma\left(3/2\right) \Gamma\left(\nu+1\right)}{\Gamma(\nu +5/2)}  \, , \label{eqApp:four}
	\end{align}
\end{subequations}
yielding 
\begin{equation}
	Z_q^{(q<1)} =  \frac{ \pi^{3/2}}{\left[\beta_q (1-q)\right] ^{3/2}} \frac{\Gamma\left(\frac{1}{1-q}+1\right)}{\Gamma \left(\frac{1}{1-q} + \frac{5}{2}\right)} \, . \label{eqApp:partition-function-q<1}
\end{equation}
We changed the variables in Eq. (\ref{eqApp:Zq-start}) defining $t = \left( 1-q \right) \beta_q R^2$, using $A =\beta_q (1-q)$, obtaining $R^2 dR = t^{1/2}/(2A^{3/2})\, dt$ and the resulting Eq. (\ref{eqApp:two}). The integral in this equation is an Euler integral of the first kind, $\int_{0}^{1} t^{x-1} (1-t)^{y-1} \, dt = B(x, y)$, leading to Eq. (\ref{eqApp:three}). The Beta function $B(x,y)$ can be expressed in terms of the Gamma function as $B(x,y) = \Gamma(x) \Gamma(y)/\Gamma(x+y)$ resulting in Eq. (\ref{eqApp:four}).\cite{Abramowitz-1964} Replacing the original variables in this last equation with $\Gamma(3/2) = \sqrt{\pi}/2$, we arrive at the final form of the partition function for the $q<1$ case, Eq. (\ref{eqApp:partition-function-q<1}). The corresponding equation for the $q> 1$ case is (supplementary material, {Sec.~ 1.2})
\begin{equation}
	Z_q^{(q>1)} =  \frac{ \pi^{3/2}}{\left[\beta_q (q-1)\right] ^{3/2}} \frac{\Gamma\left(\frac{1}{q-1} - \frac{3}{2}   \right)}{\Gamma \left(\frac{1}{q-1} \right)}\, , \quad q < \frac{5}{3} . 
	\label{eqApp:partition-function-q>1}
\end{equation}

\subsection{Normalized 3D Radial $q$-Gaussian}
\label{app:qgauss-normalized}
Replacing Eqs. (\ref{eqApp:partition-function-q<1}) and (\ref{eqApp:partition-function-q>1}) in the definition of $P_q(R)$, Eq. (\ref{eqApp:radial-q-Gaussian-Zq}),  yields the normalized $q$-Gaussian probability density 
\begin{widetext}
	\begin{subequations}
			\begin{align}
					P_q^{(q<1)}(R) &= 4\pi R^2 	\left(1-q\right)^{3/2}
					\left(\frac{\beta_q}{\pi} \right)^{3/2}  
					\frac{\Gamma\left(\frac{5}{2}+\frac{1}{1-q}\right)}
					{\Gamma\left(1+\frac{1}{1-q}\right)}
					\left(1-(1-q) \beta_q R^2\right)^{\frac{1}{1-q}}		\label{eqApp:prob-normalized-q < 1}		 \\
					P_q^{(q>1)}(R) &= 4\pi R^2 	\left(q-1\right)^{3/2}
					\left(\frac{\beta_q}{\pi} \right)^{3/2}  
					\frac{\Gamma\left(\frac{1}{q-1}\right)}
					{\Gamma\left(\frac{1}{q-1} -\frac{3}{2}\right)}
					\left(1+(q-1) \beta_q R^2\right)^{-\frac{1}{q-1}}	, \quad q<\frac{5
						}{3}\, .
					\label{eqApp:prob-normalized-q > 1}
				\end{align}
		\end{subequations}
\end{widetext}
These two equations can be expressed in the compact form of Eq. (\ref{eq:q-Gaussian-general form-radial}).

\subsection{Gaussian limit $q\to 1$}
\label{app:qgauss-limit}
For $q\to 1$ one has $\exp_q(-x)\to \exp(-x)$. To verify that the prefactor in Eqs.~(\ref{eqApp:prob-normalized-q < 1}) and (\ref{eqApp:prob-normalized-q > 1}) converge to the same Gaussian distribution, we define $\nu=1/(1-q)$ and use the asymptotic ratio of Gamma functions, $\Gamma(\nu+a)/\Gamma(\nu+b)\approx \nu^{a-b}$ for $\nu\to\infty$. For the $q<1$case $a=5/2$, $b=1$ and 
\begin{equation}\nonumber
	(1-q)^{3/2}\frac{\Gamma\!\left(\frac{5}{2}+\nu\right)}{\Gamma(1+\nu)}
=\nu^{-3/2}\frac{\Gamma(\nu+5/2)}{\Gamma(\nu+1)}
\approx \nu^{-3/2}\nu^{5/2-1}=1,
\end{equation}
showing that the $q$-Gaussian converges continuously to the Gaussian from $q<1$. A similar reasoning could be applied to the $q>1$ case.

\subsection{Second moment for $q<1$}
\label{app:qgauss-second-moment}
The second moment is defined as
\begin{equation} 
	\langle R^2 \rangle_q = \frac{I_2}{I_0} =
	\frac{\int_0^{\infty} 4\pi R^4 [\,1 - (1 - q)\beta_q R^2\,]^{1/(1 - q)} dR} 
	{\int_0^{\infty} 4\pi R^2 [\,1 - (1 - q)\beta_q R^2\,]^{1/(1 - q)} dR}\, .
	\label{eq:qgauss-second-moment}
\end{equation}
The $I_0$ was evaluated above and is given by Eq. (\ref{eqApp:partition-function-q<1}) or (\ref{eqApp:partition-function-q>1}), depending on the $q$ value.

To evaluate $I_2$ we follow the same steps as in Eqs. (\ref{eqApp:Zq-start}) to (\ref{eqApp:partition-function-q<1}). The mean-square end-to-end distances of a $q$-Gaussian random walk, expressed in terms of the Gamma functions, are
\begin{align}
	\langle R^2 \rangle_q^{(q<1)}  = \frac{3}{2\beta_q (1 - q) }   
	\frac{ \Gamma\left(\frac{5}{2}+\frac{1}{1-q}\right)}
	{ \Gamma\left(\frac{7}{2}+\frac{1}{1-q}\right)} \,  
	\label{eq:R^2-q<1}\\
	\langle R^2 \rangle_q^{(q>1)} = \frac{3} {2\beta_q (q - 1) }   
	\frac{\Gamma\left(\frac{1}{q-1} - \frac{5}{2}  \right)}
	{\Gamma\left(\frac{1}{q-1} - \frac{3}{2}  \right)}\, .
	\label{eq:R^2-q>1}
\end{align}
These equations may be simplified using the Gamma identity $\Gamma(z+1) = z\: \Gamma(z)$.
For Eqs. (\ref{eq:R^2-q<1}) and (\ref{eq:R^2-q>1}) we set  
\begin{equation}\nonumber
	z = \frac{5}{2} + \frac{1}{1-q}\, ,  \quad z = \frac{1}{q-1} - \frac{3}{2}
\end{equation}
and realize that the ratio of the two Gamma functions may be expressed as 
\begin{equation}\nonumber
	\frac{\Gamma(z)}{\Gamma(z + 1)}\, ,  \quad  \frac{\Gamma(z-1)}{\Gamma(z)}\ , 
\end{equation}
respectively. 
After algebraic simplification we obtain the same mean-square end-to-end distance for both cases $q<1$ and $q>1$, Eq. (\ref{eq:R^2-q-gaussian-final}):
\begin{equation}\nonumber
	\langle R^2 \rangle_q  = \frac{3}{\beta_q \left(7-5q\right)}\; .
	\label{eqApp:R^2-q-gaussian-final}
\end{equation}

\section{Effective Entropic Spring Constant for a $q$-Gaussian Chain}
	
To derive the entropic spring constant associated with a q-Gaussian end-to-end distance distribution, we assume that the equilibrium probability density of the end-to-end vector 
$\bm{R}$ is isotropic and given by
\begin{equation}
	P(\mathbf R) \propto \exp_q\!\left(-\beta_q R^2\right),
\end{equation}
where $\bm{R} = |\bm{R}|$ and the $q$-exponential is defined by Eqs. (\ref{eq:q-exp}) and (\ref{eq:q-exp-for-q-Gaussian-q>1}).
We restrict here attention to the case $q<1$, which is relevant for unentangled chains.

The constrained configurational free energy associated with fixing the end-to-end vector $\bm{R}$ is 
\begin{equation}
	F(\bm{R}) = -k_\text{B} T \ln P(\bm{R}) + \text{const.},
	\label{eq:free-energy-q-gaussian}
\end{equation}
yielding 
\begin{equation}
	F(\bm{R}) = -k_\text{B}T\; \ln \left[\exp_q\!\left(-\beta_q R^2\right)\right]
	  + \text{const.}
\end{equation}
For small extensions, the logarithm of the $q$-exponential can be expanded as
\begin{equation}
	\ln \exp_q(-\beta_q R^2)
	= -\beta_q R^2 + O(R^4),
\end{equation}
independently of the value of $q$.

Substituting into the free energy yields,
\begin{equation}
	F(\mathbf R) = k_\text{B}T \, \beta_q R^2 + O(R^4).
\end{equation}
The effective linear entropic spring constant is defined as the curvature of the free energy at the origin,
\begin{equation}
	k_q \equiv \left.\frac{\partial^2 F}{\partial R^2}\right|_{R=0},
\end{equation}
which gives
\begin{equation}
	k_q = 2 k_B T\, \beta_q.
	\label{eq:kq-beta}
\end{equation}

For a three-dimensional $q$-Gaussian distribution with $q<1$, the mean-square end-to-end distance is given by Eq. (\ref{eq:R^2-q-gaussian-final}), repeated below 
\begin{equation}
	\langle R^2 \rangle_q = \frac{3}{\beta_q(7-5q)}.
\end{equation}
Eliminating $\beta_q$ between this expression and Eq.~(\ref{eq:kq-beta}) yields
\begin{equation}
	k_q = \frac{6 k_B T}{(7-5q)\,\langle R^2\rangle_q},
\end{equation}
which reduces to the Gaussian result in the limit $q\! \to \! 1$.

This expression is a direct consequence of the equilibrium $q$-Gaussian statistics measured in this work. Its implications  for chain dynamics are left for future study.

\bibliography{references-q-gauss-unentangled-sorted.bib}

\end{document}